\documentclass[useAMS,usenatbib]{mn2e}

\usepackage{graphics}
\usepackage{graphicx}
\input{epsf}
\begin{document}

\def\bc{\begin{center}}
\def\ec{\end{center}}
\def\b{\begin{equation}}
\def\e{\end{equation}}
\def\ber{\begin{eqnarray}}
\def\eer{\end{eqnarray}}
\def\l{\left}
\def\r{\right}
\def\eg{{\it e.g.}}
\def \ie {{\em i.e.~~}}
\def \lleq {\lower0.9ex\hbox{ $\buildrel < \over \sim$} ~}
\def \ggeq {\lower0.9ex\hbox{ $\buildrel > \over \sim$} ~}
\def \dlt {$\delta$}
\def \ffv {FF$_{\rm V}$}
\def \ffm {FF$_{\rm M}$}
\def \ffl {FF$_{\rm LC}$}
\def \L {\Lambda}
\def \T {\tau}
\def \Gm{$\Gamma$}
\title[Morphology of Mock SDSS Catalogues]{Morphology of Mock SDSS
  Catalogues}

\author[J. V. Sheth]{Jatush V.Sheth$^{1,2}$ \\
  $^{1}$ Inter University Centre for Astronomy $\&$ Astrophysics, Pune, India \\
  $^{2}$jvs@iucaa.ernet.in \\
  } \maketitle
\begin{abstract}
  We measure the geometry, topology and morphology of the
  superclusters in mock SDSS catalogues prepared and reported by Cole
  et al.(1998). The mock catalogues refer to $\T$CDM and $\L$CDM {\em
    flat} cosmological models and are populated by galaxies so that
  these act as biased tracers of mass, conforming with the observed
  two-point correlation function measured using APM catalogue on
  scales between 1 to 10 h$^{-1}$Mpc. We compute the Minkowski
  Functionals (hereafter, MFs) for the cosmic density fields using
  SURFGEN (Sheth et al.2003) and use the available 10 realizations of
  $\T$CDM to study the effect of cosmic variance in estimation of MFs
  and Shapefinders; the statistics derived from MFs, and used to study
  the sizes and shapes of the superclusters. The MFs and Shapefinders
  are found to be extremely well constrained statistics, useful in
  assessing the effect of higher order correlation functions on the
  clustering of galaxy-distribution. We show that though all the mock
  catalogues of galaxies have the same two-point correlation function
  and similar clustering amplitude, the global MFs due to $\T$CDM show
  systematically lower amplitude compared to those due to $\L$CDM, an
  indirect, but detectable effect due to nonzero, higher order
  correlation functions. This enables us to successfully distinguish
  the two models of structure formation.
  
  We further measure the characteristic thickness (T), breadth (B) and
  length (L) of the superclusters using the available 10 realizations
  of $\T$CDM.  While T$\le$B and T, B$\in$[1,17] h$^{-1}$Mpc, we find
  the top 10 superclusters to be as long as 90 h$^{-1}$Mpc, with the
  longest superclusters identified at percolation to be rare objects
  with their length as large as 150 h$^{-1}$Mpc. The dominant
  morphology of the large superclusters is found to be {\em
    filamentary}. The thickness, breadth and planarity of the
  superclusters follow well-defined distributions which are different
  for the two models.  Thus, these are found to be sensitive to the
  cosmological parameter-set and are noted to be candidate statistics
  which can compare the rival models of structure formation. Further,
  the longest structures of $\T$CDM are found to be significantly
  longer than those in $\L$CDM. Finally, mass and volume-weighted
  dimensionless Shapefinders -- Planarity and Filamentarity -- are
  found to be well-constrained statistics useful to discriminate the
  two models.  We note some interesting effects of bias and stress the
  importance of incorporating realistic treatment of bias in preparing
  and analysing the mock catalogues.
\end{abstract}


\section{Introduction}
The present era marks a remarkable milestone in the history of
Cosmology due to an alround increase in the extragalactic database by
several orders of magnitude over last few years. These data have
improved both in quantity and in quality, due to which theoretical
models of the Universe are directly confronted with the observations,
either to be validated or to get more refined. The primordial density
fluctuations as revealed imprinted on the otherwise smooth and
isotropic microwave background radiation (MBR) have been measured with
unprecedented accuracy and resolution by WMAP \citep{bennett03}, thus
enabling us to obtain the most reliable estimates of the cosmological
parameters and confirming that the geometry of the Universe is indeed
{\em flat}.  An independent constraint from the study of extragalactic
distances using type-Ia supernovae seems to indicate that we live in a
low matter-density, cosmological constant dominated Universe, with
$\Omega_{\L}\approx$0.7. A variety of observations like, the power
spectrum study of the MBR, the clustering properties of the
galaxy-distribution and the rotational velocity measurements in
normal-size galaxies as well as in the clusters of galaxies
\citep{kneib03}, all seem to indicate that most of the remaining
fraction of the matter-density is in the form of dark matter which is
nonrelativistic, i.e., {\it cold} in nature. Thus, the cosmological
constant dominated, {\bf C}old {\bf D}ark {\bf M}atter model ($\L$CDM)
seems to be the best fitting cosmological model of the Universe that
we inhabit.  However, several anamolies have surfaced which cannot be
easily reconciled with the hypothesis of cold dark matter. For
example, an order-of-magnitude lower than predicted number density of
dwarf or low surface brightness galaxies observed in the local
Universe, the mismatch between the predicted and the observed density
profiles of these galaxies, the lack of sufficient substructure in
normal size galaxies, going against the standard hierarchical model of
structure formation, etc. are not readily explained within the
standard CDM paradigm and secondary physical processes are sought to
explain many of these anomalies.  Hence, the issues related to
structure formation and background cosmology are likely to have a
vigorous interplay with the physics of galaxy formation, a process
which may lead to deeper insights into our ideas about the
cosmological model(s) of the Universe. It suffices, hence, to devise
algorithms and statistics which enable us to compare our theoretical
predictions with a variety of observations of the Universe.

The developments on studying the microwave background radiation are
further complemented by properties of the large scale structure
revealed by increasingly deep and voluminous redshift surveys.
Indeed, redshift surveys continue to occupy a central place in our
efforts to understand the Universe for, as opposed to CMBR which gives
us a view of the early Universe, these provide us with a view of the
matter-distribution in our immediate extragalactic neighbourhood. In
addition to revealing the large-scale distribution of galaxies, a
redshift survey provides us with a statistically fair sample to
estimate many properties associated with various galaxy-populations,
e.g., the luminosity function \citep{blant03,norberg02a}, the
clustering of galaxies of various ``types'' \citep{norberg02b}, the
mass-to-light (M/L) ratio in galaxies of different morphological or
spectral types (for example, Brainerd \& Specian 2003), etc.  Further,
with the help of efficient algorithms, it is possible to identify
groups and clusters of galaxies and assess the role which their
environments play in the evolution of their constituent galaxies
{\footnote{See \cite{einasto03} for a study of the clusters and
    superclusters in the SDSS Early Data Release.}}. Such findings
provide crucial feedback to our efforts in understanding the physics
of galaxy formation. For a detailed review highlighting potential
applications and importance of redshift surveys in the study of the
Universe, see a recent review-article by \cite{lahav-suto03}.

Whereas the paradigm of structure formation due to gravitational
instability has been explored to a great extent, both analytically and
using N-body simulations, these studies have mainly focussed on the
dynamics of the dark matter. However, we now have richer quality and
quantity of data with which to confront our theoretical models. For
example, the 2 degree field galaxy redshift survey (2dFGRS) containing
about 250 000 galaxies and covering about 1500 square degree of the
sky is complete (Colless et al.2003), while the Sloan Digital Sky
Survey (SDSS) team has recently made its first data release (FDR)
reporting the redshifts of about 180 000 galaxies \citep{abaz03}.
When complete, the SDSS will cover about a quarter of the full sky and
reveal the distribution of a million galaxies upto 600 h$^{-1}$ Mpc, a
depth comparable to that attained by 2dFGRS.

In light of the above developments, the attention of the cosmology
community has steadily shifted to simulating the formation of galaxies
in dynamically evolving dark matter background. In this context, it is
worth mentioning two widely adopted approaches -- the semi-analytic
galaxy formation methods \citep{cole00,benson02} and SPH methods
\citep{weinberg02}.  However, these two approaches do have their own
limitations (Berlind et al.2003), and important insights have been
obtained by a complementary approach based on the Conditional
Luminosity Function (see Yang, Mo, van den Bosch \& Jing 2003 and a
series of earlier papers by this group).

The catalogues of ``galaxies'' prepared using the methods advocated by
these teams are to be tested against the large body of observations
due to deep, upcoming redshift surveys. Such a comparison should have
two main aims:
\begin{enumerate}
\item To employ various population statistics which reflect the
  structural properties of the ensemble of galaxies. These are mainly
  the kinematic measures defined for a given sample of a large number
  of galaxies. The luminosity function, the mass-to-light ratios of
  galaxies of various morphological or spectral types, mass-to-light
  ratios of groups/clusters, the morphology-density relation etc. form
  this class of measures and are chiefly used to validate/invalidate a
  mock catalogue.
\item To assess the large-scale distribution of galaxies and to
  quantify the clustering of the ensemble of galaxies. The two-point
  correlation function, the three-point correlation function, the
  pair-wise velocity dispersion, etc. are the most widely used
  statistics belonging to this category. These statistics reflect the
  dynamical history and clustering properties of the
  matter-distribution.
  
\end{enumerate}

The list of statistics which fall in the latter category is, however,
not complete. The statistical properties of the primordial gaussian
random phase density field are fully quantified in terms of its power
spectrum, or its Fourier counterpart, the two-point correlation
function. However, a subsequent gravitational dynamics introduces
phase correlations among the neighbouring density modes of this
density field (Sahni \& Coles, 1995). This is accompanied by a
transfer of power from larger scales to smaller scales
\citep{hamilton91} as a result of which the entire hierarchy of
correlation functions becomes nonzero.  Hence, a full quantification
of the large scale cosmic density field at today's epoch would ideally
require us to have knowledge of all the higher order correlation
functions.  Under appropriate assumptions, analytic treatment goes as
far as predicting the effect of gravitational dynamics on the 3-point
correlation function, or at best the 4-point correlation function, but
not beyond. It is to be noted that even this treatment is within the
perturbative regime and breaks down when the density contrast exceeds
unity.  However, we note that a mock catalogue could not be
satisfactorily compared with the real cosmic density field merely by a
subset of correlation functions for which analytic predictions are
available. In fact, most of the N-body simulations are
structure-normalised so as to reproduce the current rich cluster
abundance observed in our neighbourhood. In addition, mock catalogues
constructed from N$-$body simulations are usually {\it required} to
reproduce the observed two-point galaxy-galaxy correlation function
\citep{cole2df98,yang03}.  Thus, the mock catalogues of various
cosmological model(s) agree with the large scale structure (hereafter,
LSS) of the Universe {\it by construction}, and cannot be
discriminated from one another or from the observed LSS of the
Universe on the sole basis of the 2-point correlation function. We
might add that higher order correlation functions (beyond, say, the
three-point correlation function) are cumbersome to estimate for a
system comprising of large number of particles.  Furthermore, the
correlation functions offer us little intuitive insight to guide us in
interpreting the results.  Thus, distinctly different statistics are
needed to give us an handle on the effect of higher order correlation
functions. It is desirable that these statistics simultaneously offer
us an intuitively clear interpretation as well as allow us for their
numerically robust estimation.

In this paper, we propose to use {\it one} such class of measures --
the Minkowski Functionals (hereafter, MFs) \citep{meckwag94,sheth03}
and related morphological statistics, the Shapefinders \citep{sss98}
-- to quantify and characterise the large scale distribution of
galaxies{\footnote {See also, Doroshkevich et al.2003 for a
    morphological study of the SDSS (EDR) using minimal spanning
    trees.}}. Our main motivation in using the MFs for this purpose is
that (1) MFs have been shown to be dependent on the entire hierarchy
of correlation functions \citep{meckwag94,schmlz99}, and (2) the MFs
in both 2$-$D and in 3$-$D have physically important interpretation,
namely that out of ($n+$1) MFs available in $n-$D, the first $n$ of
them characterise the geometry of the excursion set under question
(these could be superclusters or voids of the cosmic web, for example)
and the remaining MF is sensitive to the topology or the connectedness
of the associated hypersurface. In recent times, there is even more
motivation for employing MFs for the quantification of LSS after
Matsubara (2003) derived the MFs for cosmic density fields in the
weakly nonlinear regime using 2$^{\rm nd}$ order Euler perturbative
expansion. With the availability of large datasets like 2dFGRS and
SDSS, it should be possible to test these predictions by working with
the cosmic density fields smoothed at sufficiently large length-scales
{\footnote {In fact, one could attack long lasting questions related
    to bias on large length-scales (of about a few tens of Mpc)
    employing a completely different line of approach of estimating
    MFs for cosmic density fields smoothed on, say a few ten's of Mpc
    and comparing the results with their theoretical predictions for
    the dark matter. See \cite{hikage03b} for an application involving
    genus.}}.

Various attempts have been made to quantify the cosmic web using MFs
\citep{bharad,sheth03,schmlzpap99}. The efficacy of MFs in quantifying
cosmic density fields as well as in making a morphological survey of
the structural elements of the cosmic web -- superclusters -- has
thereby been abundantly demonstrated. Recently, \cite{hikage03a}
provided Minkowski Functionals for SDSS (EDR).  Further,
\cite{basil03} used Shapefinders to study the morphology of the
superclusters in SDSS(EDR) and found the dominant morphology of the
superclusters to be filamentary. In the mean time, important advances
have been made concerning techniques to calculate the MFs. The
grid-based techniques based on the Crofton's formulae or the
Koenderink invariants \citep{Krofton97} have been recently
complemented by a powerful, triangulation based surface-generating
method which self-consistently generates isodensity contours for a
density field defined on a grid.  This method also makes possible an
online computation of the MFs. The method, the associated software
SURFGEN, its accuracy and application to a class of N$-$body
simulations were reported in a recent paper by \cite{sheth03}.

In this paper, we analyse the LSS in the SDSS mock catalogues
corresponding to two models of structure formation -- $\L$CDM and
$\tau$CDM. These mock catalogues were prepared by Cole et al.(1998)
(hereafter, CHWF) based on various physically motivated biasing
schemes represented in terms of analytic functions. The biasing
schemes were used to prescribe the ``formation-sites'' of the galaxies
for a given cosmological density field. Despite being superseded by
semi-analytic and Conditional Luminosity Function based techniques to
develop more realistic mock catalogues, {\em till date} these
catalogues have several advantages to their credit. This is because,
these catalogues fulfill several kinematic as well as dynamical
constraints. For example, the survey geometry and the b$_J$ band
luminosity functions for both 2dFGRS and SDSS have been incorporated,
ensuring an accurate reproduction of the expected radial selection
function in both the cases {\footnote{ We note however, that the SDSS
    redshifts are obtained using self-calibrated multi-band digital
    photometry, with the spectra obtained mainly in the Gunn-$r$ band.
    Blanton et al.(2003) have obtained the luminosity function in all
    the bands using about 140 000 galaxies from SDSS.}}.  In addition,
these catalogues mimic the survey geometry of both 2dFGRS and SDSS,
thus pausing the associated challenges in the subsequent analysis of
the LSS, and preparing one for the analysis of the actual survey data.
Further, the mock catalogues due to {\em all} the models and due to
{\em all} the various biasing schemes were constrained to reproduce
the two-point correlation function \citep{baugh96} and the power
spectrum \citep{bauef93} derived from the APM catalogue. The library
of catalogues is impressive, both due to a wide coverage of the
cosmological models which it offers, as well as in providing data due
to a variety of biasing schemes, all consistent with the actual LSS as
measured by the 2-point correlation function on scales between 1 to 10
h$^{-1}$Mpc.  Further, the library provides 10 realizations for the
$\tau$CDM model, which could be useful in assessing the effect of
cosmic variance on the statistics used to quantify the LSS. Due to
these points, these catalogues are ideal to test various statistics
which go beyond the usage of 2-point correlation function and which
are likely to be sensitive to the higher order correlation functions.
In fact, the sensitivity of all such statistics against various
biasing schemes could also be tested.

While 2dFGRS is complete and publically available and the SDSS team is
making its releases annually, we take our present investigation in the
spirit of ``{\it what to look for?}'' when dealing with these
datasets. Sheth et al.(2003) studied the geometry, topology and
morphology of the dark matter distributions in SCDM, $\tau$CDM and
$\Lambda$CDM models simulated by the Virgo group. One of the
motivations in carrying out this work is also to analyse as to what
changes are brought in the geometry and topology of the excursion sets
of $\tau$CDM and $\Lambda$CDM due to biasing{\footnote{It should be
    noted that the parameters for these models used by CHWF match with
    the models simulated by the Virgo group.}}.  By studying the
constrained ``galaxy distributions'' in conic volumes of the mock
catalogues, we further hope to confront at least, some of the
complications which one might anticipate in dealing with actual data.
The present exercise will provide us with a theoretical framework to
compare the full SDSS data sets when these become available.  

Our present study, in its scope, is similar to that carried out by
\cite{colley00}. However, unlike these workers, who studied the
topology of the mock SDSS catalogues, our thrust is to test more
complete set of statistics against the mock data. We shall of course
be using more advanced and latest calculational tools for the purpose
of our analysis.

The remainder of this paper is organised as follows. Section 2 is
devoted to a brief summary of the simulations which were used by CHWF
to generate the mock catalogues. Here we also describe a subset of the
mock catalogues which {\it we} use in our work. In Section 3, we
elaborate upon the method used to extracting the volume limited
samples from these mock catalogues for our further analysis. In
Section 4 we study the percolation properties of the samples. Section
5 briefly summarises the definition of MFs and the triangulation
method used to model isodensity contours and for calculating the MFs
for these contours.  Section 6 outlines our method of analysis and
presents our main results. We conclude in Section 7 by discussing the
implications of this work and outlining its future scope.

\section{The SDSS Mock Catalogues}
\subsection{Cosmological Models and their Simulations}
Cole, Hatton, Weinberg and Frenk (1998) (CHWF) prepared a
comprehensive set of mock catalogues within a variety of theoretically
motivated models of structure formation.  Each of the models was
completely specified in terms of a set of parameters defining the
background cosmology and a set of parameters fixing the power spectrum
of the initial density fluctuations.

The simulated models naturally fall into two categories: (1) The
$COBE$ normalised models, where the amplitude of the power spectrum is
set by the amplitude of the fluctuations measured in CMBR by $COBE$
and extrapolated to smaller scales using standard assumptions. (2) The
``{\it structure-normalised}'' models wherein the amplitude of
fluctuations is quantified in terms of $\sigma_8$, the rms density
fluctuations within spheres of radius 8 h$^{-1}$Mpc. $\sigma_8$ has
been set from the abundance of rich clusters in our local Universe.

The mock catalogues analysed in this paper are derived from the
structure-normalised models. Hence, we confine our present discussion
only to this class of models.

The description of the power spectrum is complete once the power
spectrum index $n$ and the Shape Parameter $\Gamma$ are specified. In
both the models of our intereset, $n$ = 1 (the Harrison-Zel'dovich
power spectrum), and $\Gamma$ has been chosen to be 0.25.
{\footnote{We note that the current data due to WMAP constrain the
    value of $\Gamma$ at around 0.21 which is quite close to the value
    adopted in models which we analyse in this paper.}}.  This could
be physically motivated either by having h = ${\Gamma\over\Omega_0}$
or by a change from the standard model of the present energy density
in the relativistic particles, e.g., due to a decaying neutrino model
proposed by \cite{be91}.

The $\T$CDM model is similar to standard CDM in having $\Omega_0$=1,
but it has the same power spectrum as other low-density flat models.
For the sake of testing the discriminatory power of the MF-based
statistics, we also choose to apply our methods to a realization of
$\Lambda$CDM structure-normalised {\em flat} model with
$\Omega_{\Lambda}$=0.3. The $\sigma_8$ in both the models was fixed by
the prescription $\sigma_8$=0.55$\Omega^{-0.6}_0$ as provided by
\cite{wef93}. This corresponds to $\sigma_8^{\tau}$ = 0.55 and
$\sigma_8^{\Lambda}$ = 1.13{\footnote{The current value of $\sigma_8$
    as provided by WMAP is $\sigma_8$=0.84. While our line of approach
    may remain the same, we hope to compare the observations with more
    realistic set of mock catalogues involving improved values and
    bounds on the above parameters. A detailed comparison with the
    observations is, hence, deferred to a future work.}}.

Starting with the complete specification of the model-parameters, the
initial conditions of all the simulations were set according to
glass-configuration and the simulations were evolved with a modified
version of Hugh Couchman's Adaptive PPM code \citep{couch91}. The
physical size of the box used was 346.5 h$^{-1}$Mpc and the
calculations were carried out on a 192$^3$ grid. The simulations were
evolved up to the present epoch. We refer the reader to CHWF for
further details.

\subsection{Selecting the formation sites for galaxies}
The simulations described above provide the distribution of dark
matter. In order to prescribe the preferred sites for the formation of
galaxies within a given mass-distribution, CHWF employed a set of
models which invoke a local, density-dependent bias. The rationale
behind this exercise comes from the observational evidence obtained by
\cite{peadod94}, who found the distributions of Abell clusters, radio
galaxies and optically selected galaxies to be biased relative to one
another. The biasing schemes employed by CHWF were in terms of simple
parametric functions because at the time, it was still not possible to
determine the function that relates the probability of forming a
galaxy to the properties of the mass density field. It should be noted
however, that the situation has changed in the mean time due to
development of elegant analytic tools like the Halo Distribution
Function (HDF) \citep{berlind02,kravtsov03} and Conditional Luminosity
Function (CLF)\citep{yang03} which provide us with a deeper
understanding of issues relating to bias.  {\footnote{On observational
    side, \cite{yan03} have constrained the halo model using 2dFGRS
    whereas, \cite{maglio03} study the halo distribution of 2dFGRS
    galaxies. At least in within a given cosmological model, hence, a
    robust study of bias should now be possible.}}.

CHWF ascribed the values of the parameters in the parametric,
bias-prescribing functions by constraining the distributions of
``galaxies'' within each of their cosmological simulations to
reproduce the two-point correlation function on the scales between 1
to 10 h$^{-1}$Mpc \citep{baugh96}. The $\Lambda$CDM and $\tau$CDM
models which we investigate in this paper were populated with galaxies
by following a selection probability function given by 
\begin{eqnarray}
P(\nu) &\propto& \exp(\alpha\nu + \beta\nu^{3/2}) ~~~ {\rm if}~~\nu \ge 0 \nonumber \\
       &\propto& \exp(\alpha\nu) ~~~ {\rm if}~~\nu \le 0,
\end{eqnarray}
where the dimensionless variable $\nu$ is defined to be $\nu$({\bf r})
= $\delta_{S}$({\bf r})/$\sigma_{S}$. $\delta_{S}$({\bf r}) is the
density contrast of the {\em initial} density field smoothed using a
gaussian window on a scale of 3 h$^{-1}$Mpc. $\sigma_S$ is the r.m.s.
mass-fluctuation of this field. For $\Lambda$CDM model,
$\alpha_{\Lambda}$=2.55 and $\beta_{\Lambda}$=$-$17.75. The large
negative value of $\beta_{\Lambda}$ indicates that this model requires
large anti-bias in the high density peaks of the initial density field
in order to reproduce the observed 2-point correlation function. For
the $\tau$CDM model, $\alpha_{\tau}$=1.10 and $\beta_{\tau}$=$-$0.56.
The above probability function was normalised so as to produce a total
of 128$^3$ galaxies within the cubic box of size 346.5h$^{-1}$Mpc,
which corresponds to the observed average number density of galaxies
with luminosity $\ge$ L$_{*}$/80.  Further details about this and
other biasing schemes can be found in CHWF.

\subsection{Preparing a mock catalogue}
Although CHWF have prepared mock catalogues for both 2dFGRS and for
SDSS, we shall confine our present discussion to the SDSS catalogues
alone.

When completed, the SDSS shall cover 3.11 steradian, i.e., about
1/4$^{\rm th}$ of the sky. The survey geometry of the SDSS is defined
in terms of an ellipse in the northern galactic hemisphere with the
centre of the ellipse at R.A.= 12$^{\rm hr}$20$^{\rm m}$,
$\delta$=32.8$^o$, which is in close proximity of the northern
galactic pole. The semi-minor axis of the ellipse runs for 110$^o$
along a line of constant R.A., whereas the semi-major axis subtends an
arc of 130$^o$. CHWF choose the location of an observer inside the
cube, and mimic the survey geometry around the observer's position.
The selected galaxies are taken to fall inside the survey geometry
prescribed above. The simulation cube is replicated periodically in
all the directions to achieve a depth corresponding to z=0.5. For the
sake of ideal comparison, the observer's position in {\em all} the
mock catalogues is chosen to be the same.

The SDSS operates on its own multi-band digital photometry and primary
selection of galaxies is made in Gunn-r band. However, for the sake of
simplicity, CHWF prepare the mock SDSS catalogues using a b$_J$ band
luminosity function with a Schecter form. By doing so, they are able
to reproduce the radial selection function and the expected number
count for the SDSS {\footnote{The b$_J$ band luminosity of a galaxy is
    related to the SDSS g' and r' bands: b$_J$ =
    g'+0.155+0.152(g'$-$r').}}.
The magnitude limit employed is b$_J<$18.9, so as to approximately
match the expected SDSS count of 900 000 galaxies in the survey area.
The Schecter luminosity function, which gives the comoving number density
of galaxies with luminosity between L and L+dL, is given by
\b \phi(L)dL =
{\phi_*\over L_*}\left({L\over L_*}\right)^{\alpha_*}\exp(-L/L_*)dL,
\e 
where CHWF employ $\phi_*$ = 1.4$\times$10$^{-2}$h$^3$Mpc$^{-3}$,
$\alpha_*$ = $-$0.97 and M$_{b_J}^*-$5log$_{10}$h = $-$19.5.

The luminosity function and the magnitude limit completely specify the
radial selection function, which prescribes the number of {\em
observable} galaxies as a function of radial distance r from the the
observer (assuming isotropy and no evolution in the luminosity
function over the redshift-range of interest; z$<$0.5). This
information is used to select an appropriate number of galaxies in a
shell of thickness dr at a distance r. These galaxies are further
randomly assigned luminosities consistent with the Schecter form of
the luminosity function.  Having ascribed the luminosity, CHWF also
tag every galaxy with the maximum redshift z$_{\rm max}$ up to which it
would be detectable in the survey. As it turns out, the provision of
this number makes it easier for the user to generate a volume limited
sample out of a given mock catalogue.

\section{Data Reduction: Extracting the mock samples}
\subsection{Preparing Volume Limited Samples}
With this section, we return to the subject of our present
investigation.  Our interest lies in utilising the geometric and
topological characteristics of the excursion sets of a cosmic density
field to test whether we can compare and distinguish between the rival
models of structure formation. For this purpose, we work with 10
realizations based on the mock SDSS catalogue due to $\tau$CDM model and
compare it with a realization of the $\Lambda$CDM model.

Since the excursion sets of a density field are extended objects, one
should ensure that no user-specific bias is introduced while studying
them. However, precisely such bias is encountered in dealing with {\em
  flux-limited} samples. Such samples are characterised by a selection
function which falls off upon moving radially outwards. This is
because the galaxies become fainter with increasing distance, and the
cut-off luminosity L$_{min}$(z) corresponding to the apparent
magnitude limit of the survey increases with redshift, leading to a
drop in the number of detectable galaxies.

To illustrate our point consider the case of an extended filamentary
supercluster (which could contain up to $10^{4-5}$ galaxies) and which
is extending radially outward from the location of the observer. In a
flux limited sample the selection function will ensure that galaxies
of a given brightness belonging to the rear end of the supercluster
are systematically suppressed in number compared to similar galaxies
closer by. Thus the size and shape of the supercluster will be grossly
distorted in a flux limited sample. It is clear from here that the
{\em physical definition} of a supercluster here depends on the
position and orientation of the supercluster relative to the observer.
In principle, no such dependence should be allowed in defining the
supercluster.
  
To avoid such problems and to ensure that our analysis is bias-free,
we shall first construct volume limited subsamples from the flux
limited mock surveys and then apply our methods to derive
morphological and geometrical properties of these volume limited
subsamples.
  
This is achieved by discarding all such galaxies at a given distance
r, which fall below the detection limit at the farthest end of the
sample. In turn, only such galaxies are selected from the flux-limited
sample, which would be visible throughout the volume of the sample
being analysed.  This is equivalent to rendering the selection
function spatially invariant.
  
The selection function can be written as
\b S(d_L) \propto \pi d_L^2 \Delta(d_L) \times
\int_{L_{min}}^{\infty}\phi(L)dL, 
\e
where d$_L$ is the luminosity distance corresponding to $\L$CDM and
$\T$CDM, where appropriate. The quantity L$_{min}$(z) corresponds to
the apparent magnitude limit of the survey and is an increasing
function of z.
\begin{figure*}
        \centering
        \centerline{
        \includegraphics[scale=0.9,trim=20 140 5 140,clip]{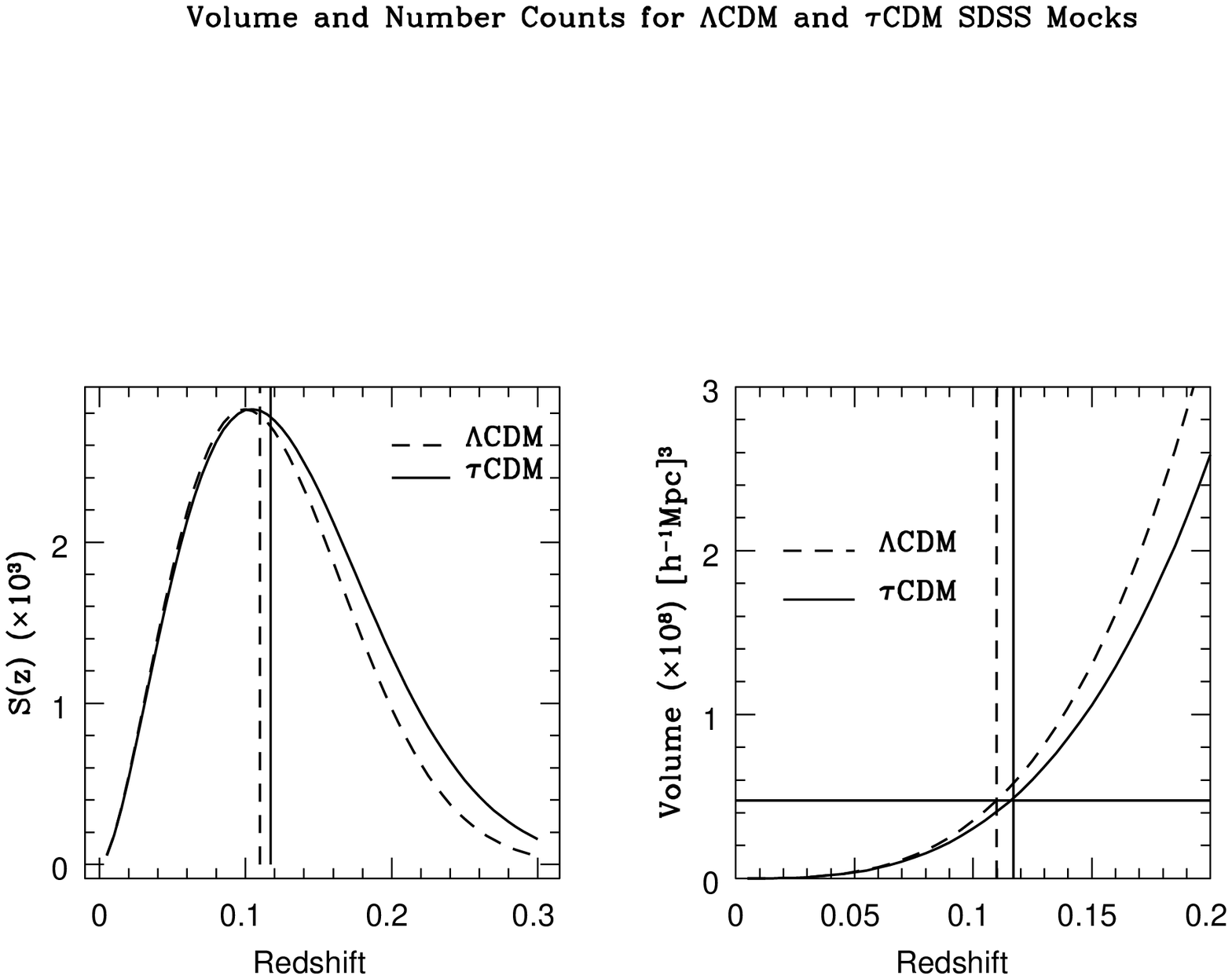}
        }
\caption{{\em Left panel}: The selection function is shown for both the 
  models. S(z) peaks around z=0.11 in both the cases and falls off on
  either side.  The solid vertical line refers to the limiting
  redshift (z=0.117) for the $\T$CDM volume limited samples. The
  dashed vertical line signifies the limiting redshift (z=0.11) for
  the $\L$CDM model. All the samples include $\sim$2.19$\times$10$^5$
  galaxies, which corresponds to a mean inter-galactic separation of
  $\sim$6 h$^{-1}$Mpc. {\em Right panel}: The survey-volume is plotted
  as a function of the limiting redshift z. We notice that the
  limiting redshifts (solid and dashed vertical lines) refer to
  similar values for the survey-volume.}
\label{fig:selfunc}
\end{figure*}
Figure \ref{fig:selfunc} shows the selection function S(z) along with
the survey-volume in (h$^{-1}$Mpc)$^{3}$ for both $\L$CDM and $\T$CDM
models. The selection function S(z) stands for the number of galaxies
included in the survey within a spherical shell of radius d$_L$(z) and
an infinitesimal thickness $\Delta$d$_L$. In both models S(z) peaks at
a redshift $\sim$0.11 and falls off on either side. The fall-off of
S(z) on the left side is a {\it volume effect}, i.e., though the
number density of detectable galaxies is higher (than at z$\sim$0.11),
the available survey-volume is small. For z$>$0.11, the number density
of galaxies is lower than at z=0.11, even though the physical volume
is large. From the point of view of analysis, one would prefer as large
a volume as possible, and a {\em constant} number density of galaxies
to pick out from all parts of this sample-volume.

Choosing the depth corresponding to z$>$0.11 increases the survey
volume at the cost of the number density.  The mean inter-galactic
separation $\bar{\lambda}$($\sim\bar{n}^{-1/3}$) in such sample(s)
will be larger. Hence, to obtain a continuous density field, we need
to smooth the galaxy-distribution more. For z$<$0.11, the
inter-galactic separation is reduced; we need to smooth less, but this
happens at the cost of the survey volume. Hence, a situation may be
envisaged wherein the structures which we study do not represent the
large scale distribution of matter in a fair manner. Thus, the peak of
S(z) signifies a compromise between two competing factors, the number
density and the survey volume. By working with the {\em
  volume-limited} sample corresponding to the peak of S(z), we
optimise on the inter-galactic separation (and hence, the required
smoothing) and on the sampling of the large scale matter-distribution
(which is proportional to the available survey volume.).

From the above analysis we expect to achieve a fair sample of LSS if
we work with a volume-limited sample of depth z$\sim$0.11. Creating a
volume-limited sample out of the mock catalogues of CHWF is relatively
simpler because, in addition to supplying the luminosity and the
redshift of the galaxy, CHWF also provide the maximum redshift
z$_{max}$ upto which the galaxy will be visible. In order to prepare a
volume-limited sample with a given depth R$_{\rm max}$, we select {\em
  all} the galaxies which are nearer to us than R$_{\rm max}$, and
which have z$_{\rm max}\ge$ Z$_{\rm MAX}$, the limiting redshift of
the sample.  We study the number of galaxies included in the sample as
a function of limiting redshift Z$_{\rm MAX}$. While we observe this
number to peak at z$\sim$0.11 in both the models (Figure
\ref{fig:selfunc}), we prefer to work with $\T$CDM samples of depth
Z$_{\rm MAX}^{\T}$=0.117 (N$_g\sim$2.19$\times$10$^5$). N$_g$, the
number of galaxies in $\T$CDM, is comparable to the number of galaxies
included in $\L$CDM catalogue if Z$_{\rm MAX}^{\L}$=0.11.  The solid
and dashed vertical lines in both panels of Figure \ref{fig:selfunc}
refer to these redshifts for $\T$CDM and $\L$CDM models respectively.
Our choice of limiting redshifts is motivated by our intention of
working with {\em same} physical volumes for both the models. The
horizontal line in the right panel of this figure shows that the
survey volumes in both models are comparable. We finally have 10
$\T$CDM samples and a $\L$CDM sample, all having practically identical
volume and a similar number of galaxies. This means that the mean
inter-galactic separation is similar in all 11 samples and we can
employ the same smoothing scale in all cases. This is essential in the
wake of assessing the regularity behaviour of MFs and to compare the
two rival models at hand.

The mock catalogues provide us with redshift-space positions of the
galaxies in terms of their Cartezian coordinates ($x$,$y$,$z$).  The
Z$-$axis of the coordinate system points towards the centre of the
SDSS survey.  The longitude $l$ (=$\tan^{-1}(y/x)$) is measured
relative to the longer axis of the SDSS ellipse, whereas the latitude
$b$=90$^o$ is the centre of the SDSS survey.  Knowing ($x$, $y$, $z$),
it is straightforward to obtain the Cartezian coordinates (X, Y, Z) of
the galaxies on a distance scale. Throughout we work with
redshift-space coordinates of the galaxies {\footnote{CHWF assume that
    the ``galaxies'' in their mock catalogues share the bulk flow with
    the matter distribution, and do not attempt to model the velocity
    field of galaxies in any more detail. Their decision was at the
    time supported by the Virgo simulation studies by Jenkins et
    al.(1998).}}, and do not attempt to correct for the redshift space
distortions. This decision is partly influenced by the analysis of
Colley et al.(2000), who found the global genus-curve of the mock SDSS
catalogue to be only mildly sensitive to redshift-distortions.  Since
the redshift space distortions are directly sensitive to $\Omega_0$,
we do anticipate definitive signatures of redshift space distortions
on the sizes of the structures.  In particular the {\it fingers of
  God} effect needs to be quantified using MFs and Shapefinders. Such
an investigation is beyond the scope of the present paper and will be
taken up in a future work.  The typical box enclosing all the samples
is characterised by (X$_{min}$, Y$_{min}$,
Z$_{min}$)$\simeq$($-$322,$-$290.5,0.0) and (X$_{max}$, Y$_{max}$,
Z$_{max}$)$\simeq$(322, 290.5, 357), where all the coordinates are
measured in h$^{-1}$Mpc.
\begin{figure}
  \centering
  \centerline{
    \includegraphics[scale=0.5]{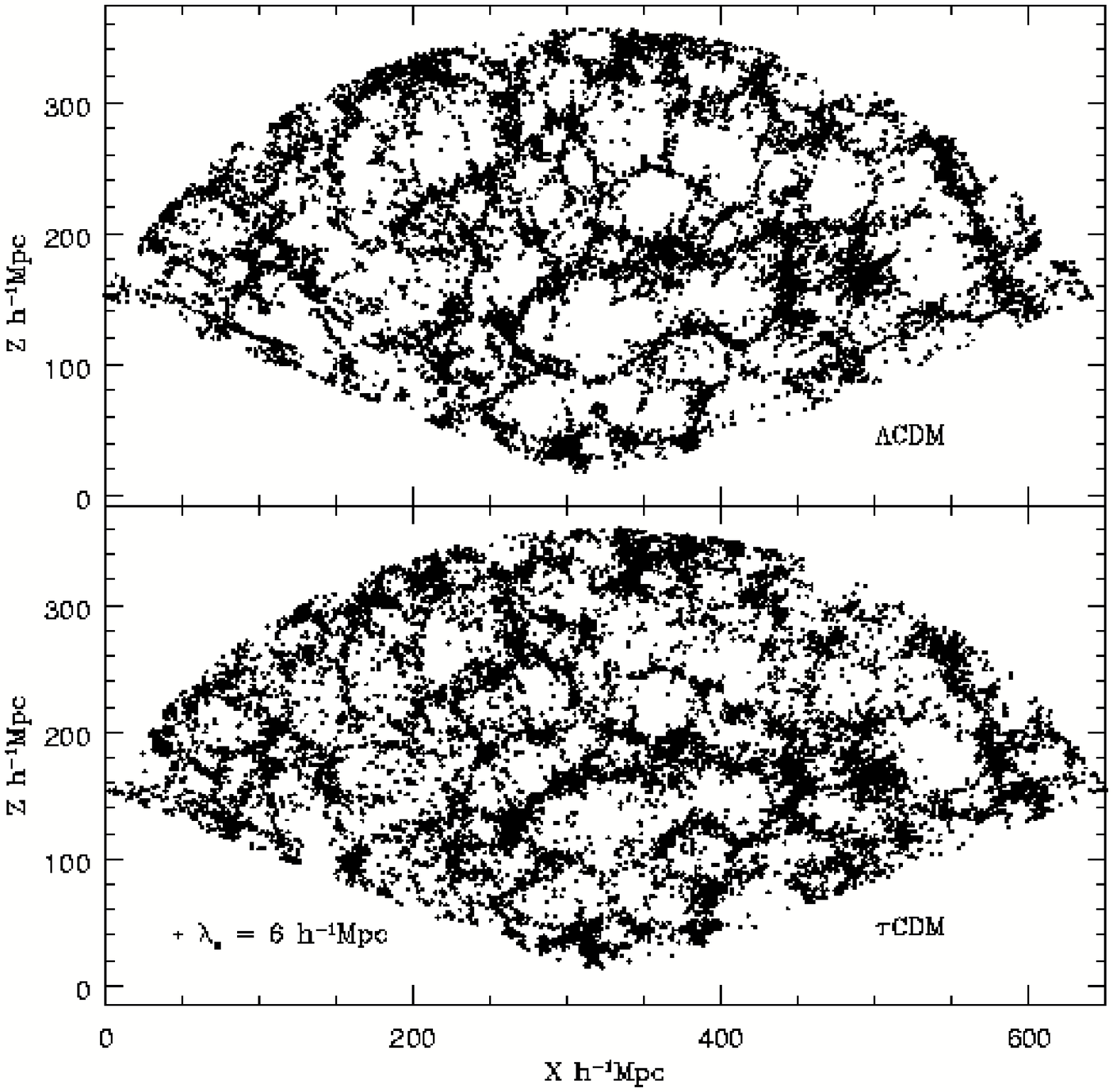}
    }
\caption{Shown here are the central slices of the $\L$CDM mock SDSS catalogue
  and of one of the realizations of its $\T$CDM counterpart. Both the
  catalogues refer to the same set of random numbers. The coordinates
  refer to the redshift space positions. The {\it fingers-of-God}
  effect is evident from these slices. The galaxy-distribution is
  constrained to match the observed 2-point correlation function, and
  the two slices qualitatively look quite similar. The voids in
  $\L$CDM are however noted to be much cleaner, and the {\em
    fingers-of-God} effect stronger. The galaxy distributions are
  smoothed with a gaussian window of size 6 h$^{-1}$Mpc. In the left
  corner of the bottom panel we show the window-size to give a feel
  for the effective smoothing done around every galaxy.}
\label{fig:slices}
\end{figure}
Figure \ref{fig:slices} shows the central slices of the $\L$CDM and
one of the realizations of $\T$CDM mock catalogues, with a similar
number of galaxies. These slices refer to the same set of random
numbers in the simulations.

\subsection{Generating Density Fields}
Our analysis will be carried out on a continuous density field
sampled on a grid with uniform resolution. Thus, having extracted the
volume limited samples, we next present our method of generating
continuous density fields for these samples.

As noted in the last subsection, the size of the box enclosing a
typical sample is 
($\Delta$X, $\Delta$Y, $\Delta$Z)$\simeq$(644, 581, 357) h$^{-1}$Mpc. We
fit a grid of size 184$\times$166$\times$102 onto this box. The size
of the cells, i.e., the resolution, turns out to be 3.5 h$^{-1}$Mpc.
About 63 per cent of the cells are found to be {\em inside} the survey
boundaries. We first carry out a Cloud-in-Cell (CIC) smoothing by
attaching suitable weights to each of the eight vertices of the cube
enclosing the galaxy.  This procedure conserves mass.  About 25 per
cent of the cells inside the survey volume acquire nonzero density
after CIC smoothing. As we noted earlier, the mean inter-galactic
separation is $\bar{\lambda}\sim\bar{n}^{-1/3}\sim$6 h$^{-1}$Mpc
$\sim$ 1.71$\times\ell_g$, where $\ell_g$ is the resolution of the
grid. Thus we do expect the CIC-smoothing to leave behind an
appreciable fraction of cells with zero density.  Ideally, a density
field should be continuous, and should be defined over the entire
sample volume. To achieve this, we need to smooth the CIC-smoothed
density field further. We employ a gaussian window
\b
\label{eq:gauss}
W_s({\bf r}) = {1\over\pi^{3/2}\lambda_s^3}\exp\l(-{|{\bf r}|^2\over\lambda_s^2}\r)
\e
for this purpose. A gaussian window, being isotropic, effectively
smooths the matter distribution within $\lambda_s$ h$^{-1}$Mpc of the
cell. Since the purpose of our present investigation is to quantify
the morphology of coherent structures, the smoothing scale $\lambda_s$
should not be very large. It should be {\em just enough} to connect
the neighbouring structures. While \cite{gott89} prescribe using
$\lambda_s\ge$2.5$\ell_g$ for such applications {\footnote{An
    isotropic smoothing with sufficiently large $\lambda_s$ (say,
    $\ge$10h$^{-1}$Mpc) will reduce our discriminatory power by
    diluting the true structures to a great extent. On the other hand,
    if one intends to compare the MFs with their theoretical
    predictions (Matsubara 2003), one should employ large scale
    smoothing in order to reach a quasi-linear regime.}}, we prefer to
work with $\lambda_s$=6 h$^{-1}$Mpc, which is comparable to the
correlation length scale in the system, and reasonably smaller than
2.5$\ell_g$. As we shall demonstrate below, the behaviour of MFs is
remarkably regular even at this smoothing scale. The lower panel of
Figure \ref{fig:slices} shows the adopted smoothing scale in the
overall distribution to develop a feel for the effective smoothing
done with the samples. It is clear from here that the large voids
should be practically devoid of matter {\em even after} smoothing and
should be accessible for morphological analysis at the lowest, sub-zero
\dlt-values, whereas the nearby clusters and superclusters should
connect up at a suitable density threshold (corresponding to
\dlt$\sim$1-2, see below).  The morphology of these superstructures
should be accessible at or near the onset of percolation. In this
paper, we shall concentrate on the shapes, sizes and geometry of such
superclusters and shall quantify these notions in more detail in
Section 6.

Prior to smoothing by a gaussian (or say, a top-hat) window, which
requires utilising FFT routines, a suitable mask for the cells outside
the survey boundaries has to be assumed. It is also to be noted that
smoothing leads to a {\it spill-over} of mass from the boundaries of
the survey to the outer peripheries. Thus, the mass {\it inside} the
survey is no longer equal to the mass contained in all galaxies. To
conserve mass, we employ the smoothing prescription of \cite{melot93}:
we prepare two samples; the first containing the CIC-smoothed density
field, with cells outside the survey volume set to zero density, and
the second {\it reference grid} wherein the cells inside the survey
volume are all set to unity, while ones outside the survey boundaries
are set to zero. Both density fields are smoothed with the gaussian
window (Eq.\ref{eq:gauss}) at $\lambda_s$ = 6h$^{-1}$Mpc {\footnote{We
    use the Fastest Fourier Transform in West (FFTW) routines for the
    fourier transforms carried out on our grid. Since the
    grid-dimensions here are not powers of 2, standard fourier
    transform routines cannot be used for this purpose.}}. The {\it
  reference grid} helps us correct for the effect of having a
smoothing window at the boundaries of the survey.  We divide the
density values of the first grid, cell-by-cell, by the entries for the
density values in the reference grid. We find that this corrects for
the {\it spill-over} of mass at the boundaries, and we indeed find
mass inside the survey volume to conserve to a very good accuracy.

Having elaborated on our method to prepare volume limited samples and
to recover the underlying density fields, we now elaborate upon our
method of analysis and finally present our results. Before embarking
on quntifying the morphology of density fields, we shall study the
percolation properties of the samples. This is the subject of the next
section.

\section{Percolation of the Large Scale Structure}
Percolation gives us useful insights into the connectivity of the
density fields. Earlier workers (for example, Sahni, Sathyaprakash \&
Shandarin 1997) have found noticeable correlations between the
topological behaviour of density fields and their percolation
properties. Density fields showing bubble topology, meatball topology
or sponge/network topology have distinct features in the way such
fields percolate.  The 3-D {\it idealized} gaussian random fields
(hereafter, GRFs), which often serve as a useful benchmark in the
study of large scale structure, percolate at 16 per cent of the
overdense/underdense volume filling fraction \citep{shandzed89}, and
those defined on a grid percolate at $\sim$30 per cent volume fraction
for a variety of power spectra \citep{ys96}.  Overdense regions
undergoing gravitational clustering percolate at lower values of the
filling fraction as compared to GRF.  As a result, the difference
between the thresholds of percolation (measured in terms of the volume
filling fraction) serves as a useful diagnostic with which to quantify
the degree of clustering. Such fields show network topology and are
characterised by ``clusters'' having larger length-scales of
coherence. Various workers (e.g., Sheth et al., 2003) also find that
the dominant morphology of the superclusters (defined for a field when
the percolation sets in) is {\em filamentary}. Thus, studying the
percolation properties of our present catalogues can help us identify
a threshold of density at which to look for {\em realistic} structures
in our survey volumes.

In the ensuing analysis, our results shall refer to 11 density fields
(10 mock catalogues due $\T$CDM and one due to $\L$CDM) scanned at 50
thresholds (or levels) of density. This sampling is not uniform. Our
preliminary analysis shows that overdense regions percolate when the
overdense volume fraction $\le$ 10 per cent of the survey volume. On
the other hand the underdense regions or {\em voids} percolate when
the overdense volume fraction is $\sim$70 per cent. Thus, from a
morphological perspective, the volume fractions $\le$30 per cent and
$\ge$70 per cent are interesting for overdense and underdense regions
respectively. In the intermediate range, both the overdense and
underdense regions percolate, and most of the volume is occupied by
the largest cluster and the largest void respectively. So practically
very little is gained by studying the structures in the range FF$_{\rm
  V}\in$(0.3,0.7), where FF$_{\rm V}$, the overdense volume fraction is
given by 
\b
\label{eq:ffv}
FF_{\rm V} = {1\over V_T}\int \Theta(\rho-\rho_{\rm TH}) d^3x.
\e
Here $\Theta(x)$ is the Heaviside Theta function. 
\begin{eqnarray}
\label{eq:theta}
\Theta(x) &=& 1 ~~~x \ge 0 \nonumber \\
          &=& 0 ~~~x < 0.
\end{eqnarray}
$FF_{\rm V}$ measures the volume-fraction in regions which satisfy the
`cluster' criterion $\rho_{\rm cluster} \geq \rho_{\rm TH}$ at a given
density threshold $\rho_{\rm TH}$. V$_T$ is the total survey volume.
In the following, we use $FF_{\rm V}$ as one of the parameters with
which to label density contours. Finally, the ranges FF$_{\rm
  V}\le$0.3 and FF$_{\rm V}\ge$0.7 each are attributed 23 levels of
density, each equi-spaced in FF$_{\rm V}$. The intermediate range
FF$_{\rm V}\in$(0.3,0.7) is sampled at four levels all equi-spaced
with $\Delta$FF=0.1.
\begin{figure*}
  \centering \centerline{ 
    \includegraphics[scale = 0.7]{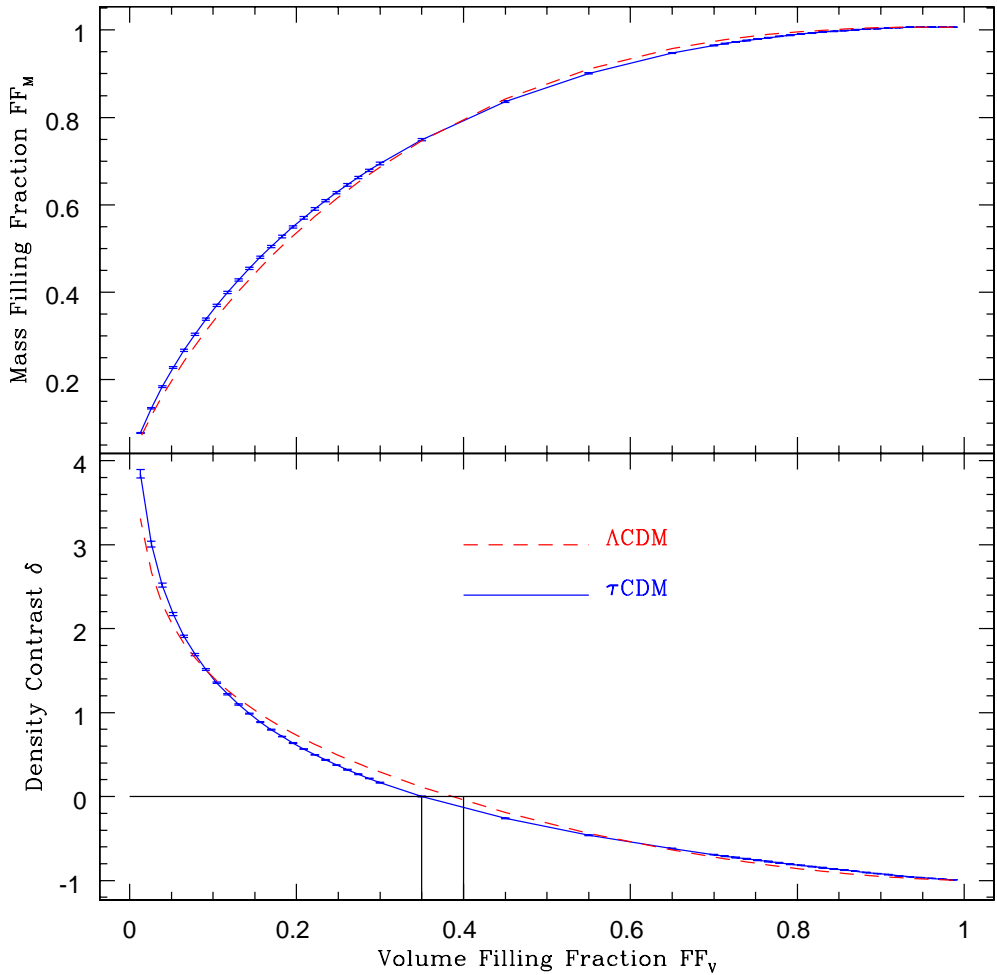} }
  \caption{The density contrast and the mass filling fraction FF$_{\rm M}$ are
    plotted as functions of FF$_{\rm V}$, the volume filling fraction
    which is treated as an independent parameter. The solid line stands
    for the $\T$CDM model and is obtained by averaging over 10
    realizations. The dashed line stands for the $\L$CDM model. Please
    refer to the text for further comments.  }
  \label{fig:ffdelm}
\end{figure*}
In Figure \ref{fig:ffdelm} we study the density contrast \dlt ~and the
mass filling fraction \ffm ~with respect to \ffv {\footnote{Throughout
    our discussion, we take the mass enclosed within a supercluster to
    be equivalent to the number of galaxies enclosed, since the two
    differ from each other only by the mass of the particle in the
    simulations, which is a constant. Though this is different for the
    two models, the mass filling fraction which usually enters our
    discussion, is unaffected by this.}}. The first noticeable effect
which the biasing brings in is, that unlike the dark matter
simulations \citep{sheth03}, \ffm ~no longer distinguishes the two
models. At all the chosen levels, the overdense mass fraction, i.e.,
the fraction of galaxies enclosed by a given normalised volume \ffv,
~is similar in $\L$CDM and $\T$CDM. The mass-parametrisation which was
noted to be so effective in distinguishing dark matter distributions
in various models, {\em cannot} perform any better than \ffv. We shall
return to this issue in Section 6. We further note that, the density
contrast \dlt ~follows similar pattern in both the models. 

We employ a {\it friends-of-friends} (FOF) based cluster-finding code
to study the number of clusters and the overdense volume fraction in
the largest cluster (\ffl) at all the 50 density-levels. \ffl ~stands
for the volume of the largest cluster normalised by the total {\em
  overdense} volume.

\begin{figure*}
        \centering \centerline{
        \includegraphics[scale=0.9]{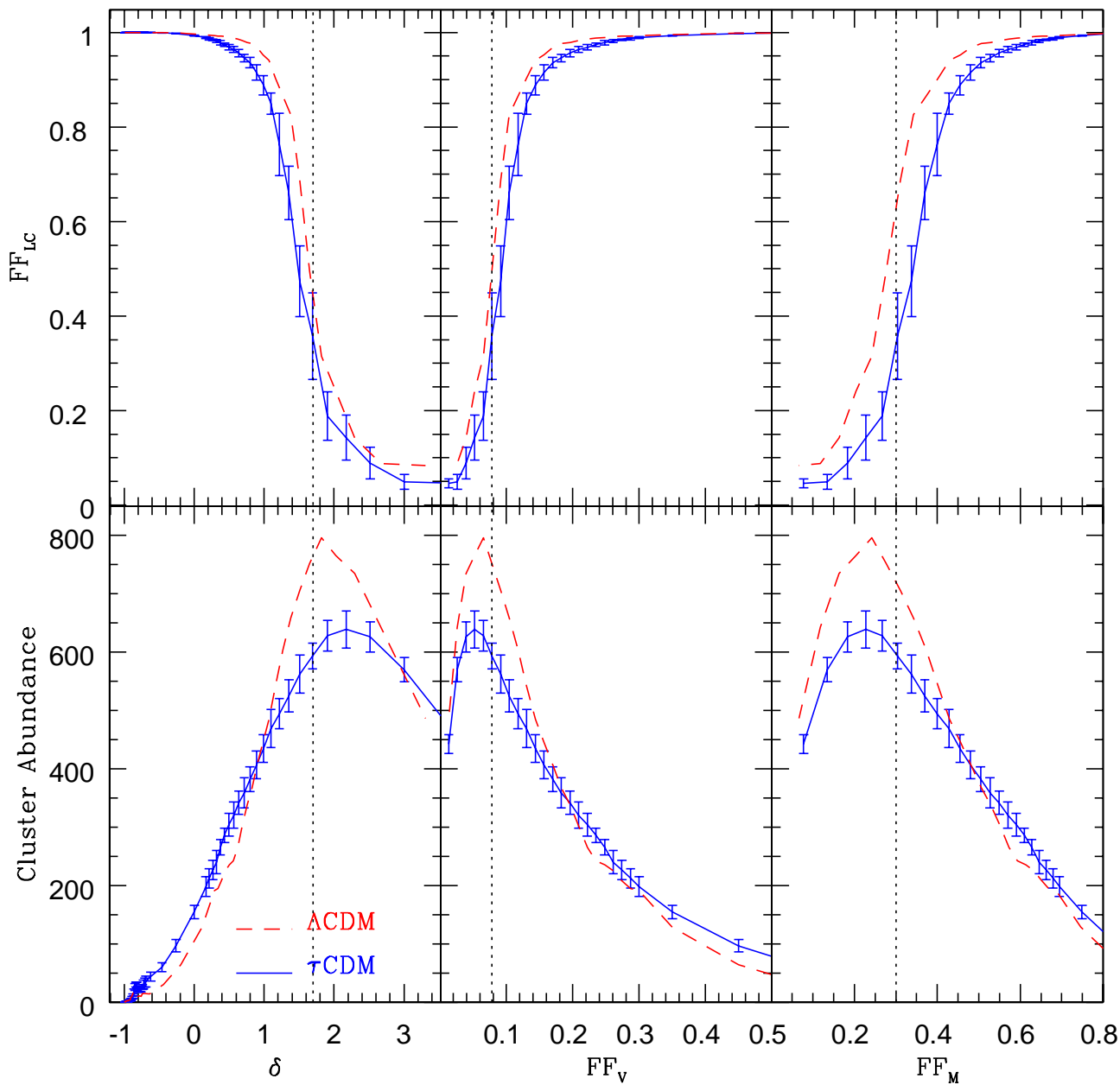} }
\caption{We study here the cluster abundance and FF$_{\rm LC}$, the fraction
  of overdense volume in the largest cluster, with three different
  parameters, $\delta$, FF$_{\rm V}$ and FF$_{\rm M}$. The $\T$CDM
  curves are averaged over 10 realizations. The percolation of the
  volume of the largest cluster is remarkably similar in both the
  models. The percolating supercluster occurs at $\delta_{\rm
    perc}^{\L,\T}$=1.7 in both $\L$CDM and $\T$CDM. The dotted
  vertical lines in all the plots signify the onset of percolation. We
  also note from the lower panels, that the number of clusters in
  $\L$CDM are larger than in $\T$CDM. This could be a useful means to
  distinguish these two models.}
\label{fig:perclt}
\end{figure*}

For cubic samples, the percolation is said to have set in when the
largest cluster first spans across the full length of the box (by
periodicity arguments, this is equivalent to having the supercluster
spanning across the infinite space). In the case of redshift surveys
wherein the matter-distribution generally falls within a cone, we
cannot utilise such a convention, for the survey boundaries diverge as
we go radially outwards and a supercluster spanning the survey volume
could as well be touching the two boundaries nearer to the observer,
without visiting other parts of the survey volume. Clearly, this does
not mean the system has percolated. Thus, for conic survey volumes,
{\em spanning the survey volume} is not a robust concept to signify
the onset of percolation. Instead, \cite{ksh93} motivate utilising the
merger property of clusters to define the onset of percolation.
Following this convention, we measure the {\em rate of growth} of the
volume of the largest cluster. The {\em highest rate of growth}
signifies pronounced merger of clusters, leading to a single, larger
cluster.  The threshold of density at which the growth-rate is highest
is defined to be the percolation threshold.
\begin{figure*}
        \centering
        \centerline{
        \includegraphics[scale=1.8,trim = 190 190 190 190]{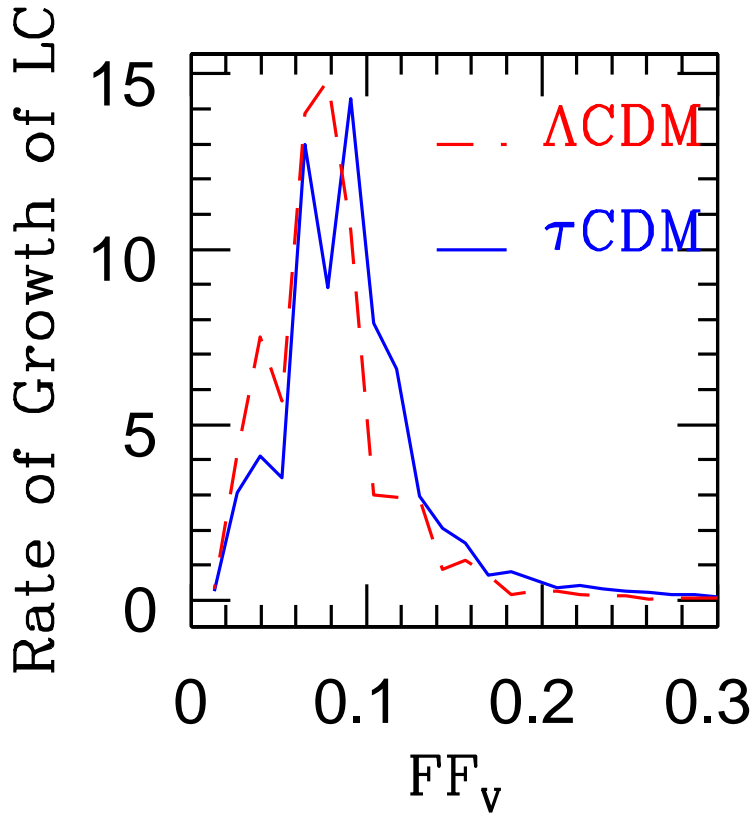}
        }
\caption{For samples with arbitrary survey geometry, the growth-rate of 
  the volume of the largest cluster is important for quantifying the
  onset of percolation. Here this quantity is plotted w.r.t. FF$_{\rm
    V}$. The merger activity in $\T$CDM is pronounced for
  \ffv$\in$[0.06,0.1]. We take FF$_{\rm V}\sim$0.08 to be the
  percolation threshold for this model. Similarly, in case of $\L$CDM,
  the largest cluster growth is highest at FF$_{\rm V}\sim$0.08.
  Hence, we take FF$_{\rm perc}^{\L}$=0.08. Where needed, we explore
  the whole range of density levels \dlt$\in$[\dlt$_{\rm
    perc}$,\dlt$_{\rm max}$].}
\label{fig:slope}
\end{figure*}
Figure \ref{fig:slope} shows the rate of growth of the largest cluster
as a function of \ffv. ~For $\T$CDM, we use the data after averaging
over all 10 mock realizations, and find that the merger rate is
highest in the range \ffv$\in$[0.06,0.1]. For our purpose, we take
\ffv ~of 8 per cent to be the percolation threshold for $\T$CDM model.
We find that the largest cluster of $\L$CDM undergoes a pronounced
merger at \ffv$\sim$0.8, bringing in 30 per cent of the overdense
volume fraction into the largest cluster.  (see Figure
\ref{fig:perclt}). We have only one realization for the $\L$CDM model,
so a statistically robust statement cannot be made, but for all
purposes, we treat \ffv = FF$_{\rm V}^{\rm perc}$=0.08 as the
percolation threshold for $\L$CDM. Where needed, we explore the entire
range of \ffv $\le$0.4 for various statistics.  To summarise,
\ffv$_{\rm ,perc}^{\L}\simeq$\ffv$_{\rm ,perc}^{\T}$=0.08.

Finally, Figure \ref{fig:perclt} shows the percolation properties of
our samples. We study the cluster abundance (lower panels) and the
filling fraction in the largest cluster \ffl ~(the upper panels) as
functions of three different parameters, \dlt, \ffv ~and \ffm. The
parameter values corresponding to percolation are marked with a dotted
line for both the models. 
For $\T$CDM, we again show the curves after averaging over the 10
available realizations. We note that the clusters are slightly
over-abundant in $\L$CDM compared to $\T$CDM.  Assuming a similar
degree of scatter in the cluster-abundance, we conclude that the {\em
  Number of Clusters} statistic is useful to discriminate the two
models. The quantities \ffv$^{\rm perc}$ are quite closer to each
other for both models as we noted earlier (middle, upper panel).  From
here, we conclude that the spatial connectedness, as measured in terms
of percolation of the volume of the LSS alone is not sufficient to
distinguish between the two models.  The top-right panel does indicate
that in $\T$CDM, 30 per cent of the total mass is in ``clusters'' when
percolation sets in, and 32 per cent of this mass (which amounts to
$\sim$22000 {\em galaxies}!) is in the percolating supercluster.
Looking at the similar volumes that these systems occupy at
percolation (although the $\L$CDM percolation sets in a little earlier
than in case of $\T$CDM), clearly other geometric and topological
descriptors like Minkowski Functionals would be more insightful to
assess the properties of the superclusters. In passing, we note two
important aspects of the models: (1) The upper left panel shows that
\dlt ~is distributed over the same range of values for both the
models.  (2) {\em For both the models, \dlt$_{\rm perc}\in$[1,2]},
which coincides with the conventional definition of superclusters.
Thus, this range in \dlt ~is in general useful for studying the
network of superclusters. We shall make use of this fact in our
subsequent analysis.

\section{Estimating Minkowski Functionals}
The work of \cite{gott86} led to a revival of interest in the paradigm
of studying and quantifying the excursion sets of the cosmic density
fields; an idea which was successfully propounded earlier in the
topological context by Doroshkevich (1970). These early efforts
however, were devoted to studying the topology and connectivity of
isodensity contours of the large scale structure (hereafter,
LSS)\citep{melotopo90}.  It was soon realised that topology {\em
  alone} may not be sufficient to describe LSS, and that it should be
complemented with geometrical information pertaining to excursion
sets. This goal was met in the form of the Minkowski Functionals
(hereafter, MFs)\citep{meckwag94}. A set of grid-based methods (e.g,
the Crofton's formulae, the Koenderink Invariants) and a method to
calculate MFs for point processes (the Boolean grain model) were
proposed \citep{meckwag94,Krofton97} to address the issue of
estimating the MFs. (In fact, the CONTOUR3D code to calculate the
topology of LSS by \citet{weinberg88} is also grid-based.)

Sheth, Sahni, Shandarin \& Sathyaprakash (2003) proposed a novel
method to calculate MFs by following a mathematically precise
prescription with which to construct isodensity contours while
quantifying them, as opposed to grid-approximations to contours
implicit in earlier treatments. Their method accurately modelled
isodensity contours in addition to performing an online calculation of
MFs for these contours.

In 3$-$D, the MFs are four in number. The first three of these are
geometric in nature, and change when we deform the surface, whereas
the fourth is a topological invariant and tells us how many handles
does a surface have in access of the holes which it completely
encloses. The four MFs are
\begin{itemize}
\item {[1]} {\it Volume} $V$ enclosed by the surface, 
\item {[2]} {\it Area} $S$ of the surface, 
\item {[3]} {\it Integrated mean curvature} $C$ of the surface,
\b
\label{eq:curv}
C = \frac{1}{2}\oint{\left({1\over R_1} + {1\over R_2}\right)dS},
\e
\item {[4]} Integrated gaussian curvature, or the Euler characteristic,
\b
\label{eq:euler} \chi = \frac{1}{2\pi}\oint{\left({1\over
      R_1R_2}\right)dS}.  
\e
\end{itemize}

A related measure of topology is the genus $G = 1 - \chi/2$, which is
what we use in all our analysis.

We shall evaluate MFs for mock catalogues by triangulating isodensity
surfaces using SURFGEN. SURFGEN is a fusion of a robust surface
modelling scheme and an efficient algorithm to calculate the geometric
quantities. The algorithm to calculate MFs on a triangulated surface
and the code SURFGEN have been discussed extensively by \cite{sheth03}
and we refer the reader to that paper for further details.
On a dual processor DEC ALPHA machine, SURFGEN calculates the MFs for
{\em all} the clusters found at about 100 levels of density on a
typical 128$^3$ grid in within a few hours' time. This amounts to
probing the MFs for about 100 000 clusters of varying sizes. In fact,
in our present application, which involves a grid which is 1.5 times
larger than this, the calculations for all the 11 samples could be
accomplished in within 2 days' time.  This highlights the scope of
using SURFGEN in more ambitious applications involving larger number
of realizations studied at higher resolution.  Our method of analysis
and the results are described in the next section.

\section{Geometry, Topology and Morphology of Mock Catalogues}
The mock SDSS catalogues being studied here, by construction, have the
same two-point correlation function in the range 1 to 10 h$^{-1}$Mpc. 
Since Minkowski Functionals are shown to depend on the entire
hierarchy of correlation functions, they may be expected to
differentiate between models as similar as $\T$CDM and $\L$CDM. This
will indeed be the case, as we shall show below. 

One of the fascinating aspects of the LSS is related to its visual
impression.  The LSS as revealed in various dark matter simulations
and redshift survey slices reveals a {\em cosmic web} of
interconnected filaments running across the sample, separated by
large, almost empty regions, called {\em voids} (see Figure
\ref{fig:slices} for an illustration). Our study will focus both on
global MFs which characterise the full excursion sets of the cosmic
density field and partial MFs which, together with Shapefinders, help
us study the shapes and sizes of the structural elements
(superclusters) of the cosmic web {\footnote{We do not however attempt
    to study the morphology of voids, which is an equally important
    constituent of the cosmic web.}}.

We begin by first studying the global Minkowski functionals. Next we
focus on how the largest cluster evolves with the volume fraction. The
last two subsections deal with the morphology of individual
superclusters in $\T$CDM and $\L$CDM models.

\subsection{Global Minkowski Functionals}
As mentioned earlier, we sample the 11 density fields (1 $\L$CDM + 10
$\T$CDM) at 50 levels of density. The criteria adopted for their
choice has been described earlier (see Section 4). The levels
uniformly correspond to the {\em same} set of volume fractions \ffv,
~for {\em all} the samples, which makes comparison between models
easier.

We find clusters using the friends-of-friends (FOF) algorithm based
Clusterfinder code at every level of density. Next we run SURFGEN on
all the clusters and estimate their MFs by modelling their surfaces.
These are the {\em partial MFs} and are useful for studying the
morphology of individual clusters{\footnote{It should be noted however
    that the ``clusters'' referred to in the present discussion {\em
      do not} coincide with the galaxy-clusters. Our clusters are
    connected, overdense regions and at appropriate level(s) of
    density may coincide with the superclusters of galaxies.}}.  We
use the {\it additivity} property of the MFs and evaluate {\em global}
MFs by summing over the contribution of {\em partial} MFs due to all
the clusters at a given level of density.  For $\T$CDM, we use the
measurements from available 10 realizations to produce average global
MFs.
\begin{figure*}
  \centering \centerline{
    \includegraphics[scale=0.8]{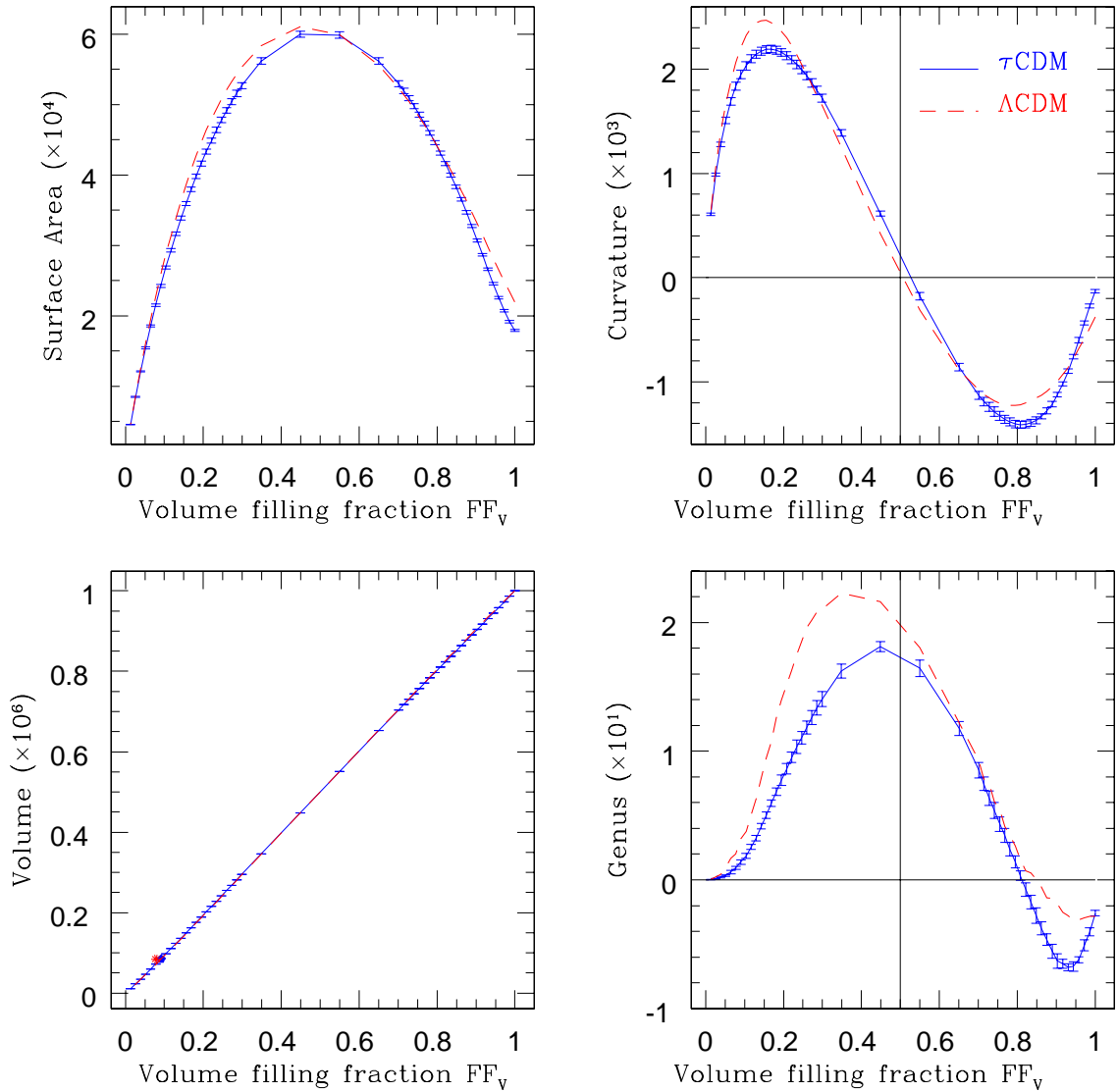} } \caption{Global
    Minkowski functionals are evaluated at 50 density levels at a
    common set of volume filling fractions and are studied here w.r.t.
    \ffv. We show the values of global MFs as normalised to the volume
    of [100 h$^{-1}$Mpc]$^3$. ~The global MFs for $\T$CDM model are
    averaged over the available 10 realizations (solid lines) and the
    errorbars represent 1$\sigma$ deviation. Assuming the same level
    of accuracy for $\L$CDM, we may conclude that the global MFs with
    volume parametrisation can indeed {\em clearly} distinguish
    $\T$CDM from $\L$CDM. Considering the fact that all the samples
    show the same two$-$point correlation function, the above
    difference can be attributed to the higher order correlation
    functions. MFs are able to respond to such a difference, and hence
    can be taken to be robust measures for quantifying the cosmic web.
    For further discussion, please refer to the text.}
  \label{fig:glov}
\end{figure*}
Figure \ref{fig:glov} shows the dependence of the global MFs on the
volume fraction \ffv. ~The solid curves represent the mean global MFs
due to $\T$CDM model and the error-bars stand for the 1$\sigma$
standard deviation. The dashed curves stand for the $\L$CDM model
{\footnote{It should be noted here that our samples {\em do not}
    exhibit periodic boundary conditions. Hence, as explained in
    detail by \cite{sheth03}, our measurements {\em cannot be}
    compared with the theoretical predictions straightaway, which are
    generally for periodic samples. In fact, the present measurements
    refer to the {\em overdense} excursion sets, and should be
    complemented by those for {\em voids}, to finally get results for
    an equivalent periodic sample. See Shandarin, Sheth \& Sahni
    (2003) for further details. The boundary effects in the present
    measurements are evident at large \ffv, ~where the area and
    curvature saturate to nonzero values referring to an
    all-pervading, cone-filling ``cluster''.}. First, we note that the
  global MFs of the $\T$CDM model are extremely well constrained. This
  shows the remarkable accuracy with which SURFGEN will enable us to
  measure the MFs of the LSS due to large datasets like SDSS and
  2dFGRS. We further note that the two models {\em can be
    distinguished} from each other with sufficient confidence using
  the global area, mean curvature and genus measurements at
  \ffv$\le$0.5 {\footnote{This is to be expected, because 50 per cent
      volume fraction refers to $\sim$85 per cent of the mass fraction
      (Figure \ref{fig:ffdelm}). Thus, the excursion sets at \ffv
      $\le$0.5 already contain most of the mass, and the necessary
      geometrical and topological information is conveyed by studying
      them.}}, whereas the global volume scales linearly with \ffv,
  ~as expected. Distinctly different trends of MFs in two models are
  due to the contribution of the higher order correlation functions,
  and do indeed reflect the manner in which the galaxies cluster in
  $\L$CDM as against $\T$CDM. This can be dubbed to be the success of
  MFs as robust statistics ideal for quantifying the cosmic web.
  
  From the dynamical point of view, these results carry some
  surprises. As noted by \cite{melotopo90,springtop}, the amplitude of
  the genus-curve drops as the N$-$body system develops phase
  correlations. Given two density fields, the system with larger
  genus-amplitude shows many more tunnels/voids which are, therefore,
  smaller in size. As time progresses, the voids are expected to
  expand and merge, leading to a drop in the genus-amplitude, while
  the phase correlations continue to grow. With this simple model, one
  could correlate the amount of clustering with the relative {\it
    smallness} of the amplitude of genus, and therefore, of the area
  and the mean curvature. As illustrated by \cite{sheth03}, the dark
  matter distribution of $\T$CDM model due to Virgo group consistently
  shows considerably larger amplitudes for the MFs compared to the
  $\L$CDM model {\footnote{However, $\Gamma_{\rm Virgo}$=0.21, which
      is different from $\Gamma$=0.25 adopted in the simulations
      studied here. So long as $\Gamma$ is the same for both the
      models, the conclusions are expected to remain unaltered.}},
  whereas, we find the reverse trend in the MFs of ``the galaxy
  distributions'' due to the same two models. This is a sufficiently
  robust result because, the $\T$CDM curves are averaged over 10
  realizations. We could interpret this as an instance of the {\em
    phase-mismatch} between the galaxy-distribution and the underlying
  dark matter distribution and could attribute it to the biasing
  prescription invoked.  This is supported by the fact that CHWF
  require large anti-bias in high density regions of the $\L$CDM
  model.
  
  To conclude, the study of the global MFs reveals from here that,
  local, density-dependent bias could lead to an apparent {\em
    phase-mismatch} or different phase-correlations among the dark
  matter and the galaxies. In passing, we also note that $\L$CDM
  preserves its unique {\em bubble shift} which is pronounced compared
  to $\T$CDM (lower right panel of Figure \ref{fig:glov}).
\begin{figure*}
  \centering
  \centerline{
    \includegraphics[scale=0.8]{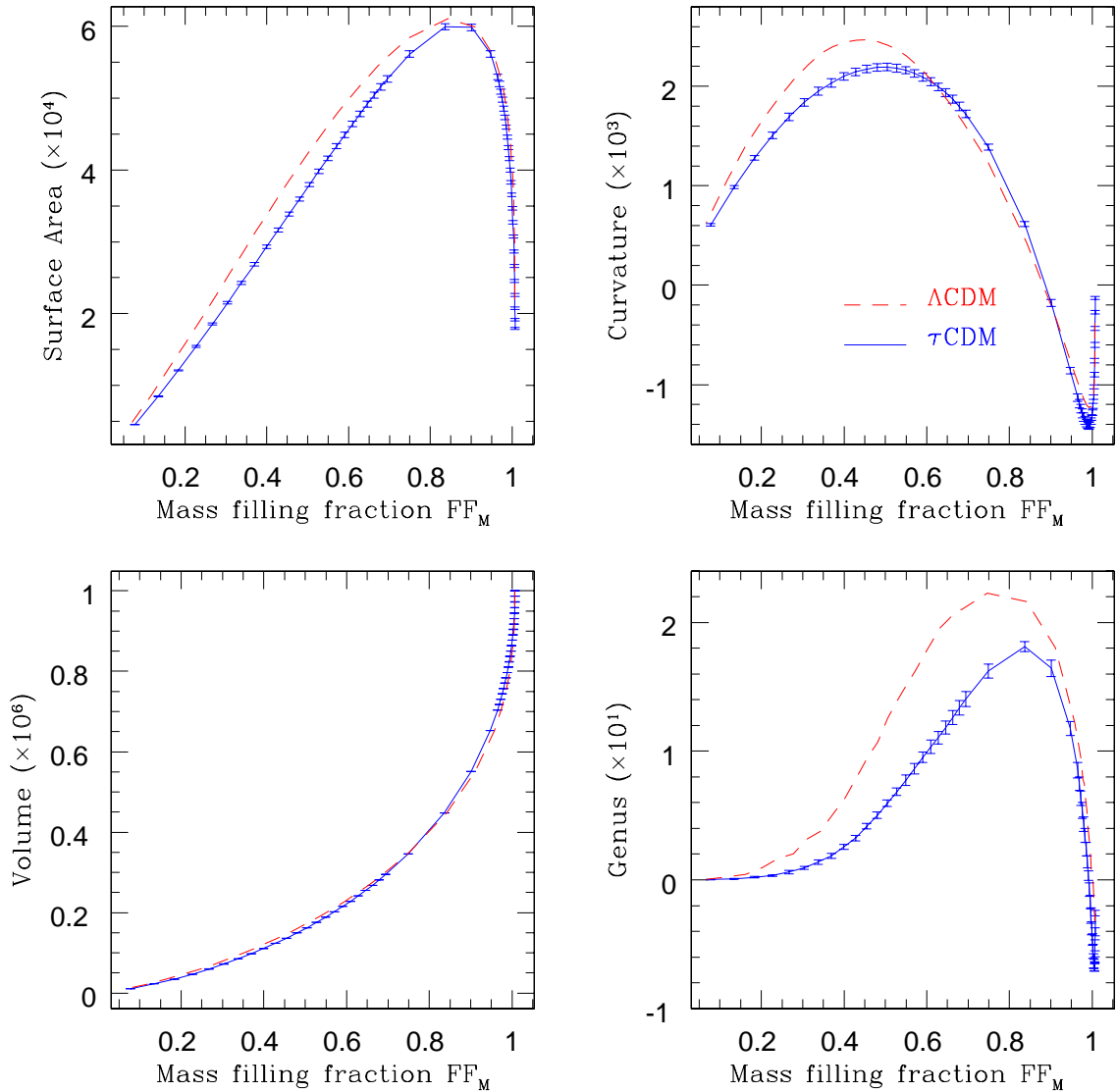}
    }
  \caption{Global Minkowski functionals are studied here w.r.t. \ffm, ~which 
    is defined as the overdense mass fraction. The MF-values are normalised 
    to a volume of [100 h$^{-1}$Mpc]$^3$. In addition to the fact
    that the models {\it can be} distinguished, we also note that the
    geometry and topology of the excursion sets, when coupled with the
    overdense mass fraction, is markedly different in both the models
    at and beyond the onset of percolation.}
  \label{fig:glom}
\end{figure*}

Figure \ref{fig:glom} shows the global MFs plotted against the
overdense mass-fraction \ffm. ~The mass-volume relation is remarkably
similar among both the models. This again is an effect of biasing. In
addition to the fact that the models {\it can be} distinguished, we
also note that the excursion sets enclosing the same fraction of mass
due to the two models show appreciably different geometry and
topology. Thus, assessing the excursion sets enclosing the same
mass-fraction could be a useful means of analysing the realistic data
and comparing them with the mock data. 

\subsection{Minkowski Functionals for the Largest Cluster}
As we scan the density fields by lowering the density threshold, the
largest cluster of the system grows in size due to merger of smaller
clusters.  The manner of evolution of the largest cluster is connected
with the percolation properties of the system. We noted earlier
(Section 4) that the largest cluster spans most of the volume by the
time the volume fraction reaches \ffv$\sim$0.3. In fact, most of the
features in the global MFs beyond \ffv=0.3 pertain to the evolution of
the largest cluster alone.  Hence, in this section, we study the
evolution of the MFs of the largest cluster for \ffv$\le$0.3. We find
that, of the four MFs, the integrated mean curvature (hereafter, IMC)
and the genus are the best discriminants.  Further, it is instructive
to see the behaviour of the largest cluster with \ffm$-$parametrisation.
In Figure \ref{fig:maxcl}, we show the IMC and genus for the largest
cluster as functions of \ffv ~(left panels) and \ffm ~(right
panels). 
\begin{figure*}
  \centering
  \centerline{
    \includegraphics[scale=0.8]{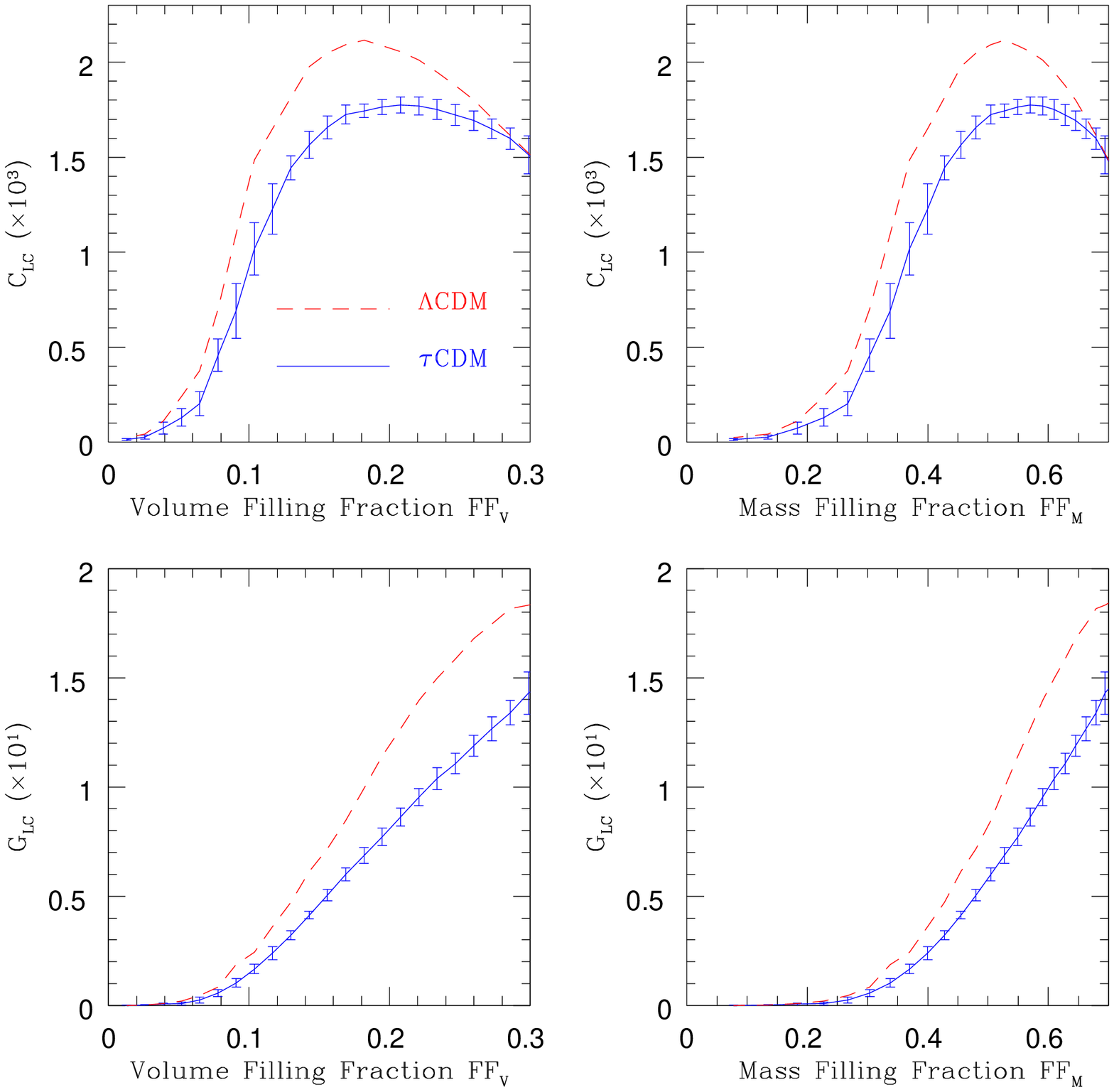}
    }
  \caption{The largest cluster spans most of the sample-volume for \ffv$>$0.3.
    Here we study the dependence of the mean curvature and genus of
    the largest cluster of $\T$CDM and $\L$CDM w.r.t. \ffv~(left
    panels) and \ffm~(right panels). Note the well-constrained
    behaviour of both the measures for $\T$CDM, and the distinctly
    different trend observed for them in $\L$CDM. Please refer to the
    text for the discussion.}
  \label{fig:maxcl}
\end{figure*}
We notice in the bottom-left panel of this figure that the genus of
the largest cluster remains vanishingly small until the onset of
percolation in both the models.  However, once percolation sets in,
the genus of the largest cluster grows much more rapidly in $\L$CDM
than in $\T$CDM. The larger genus is indicative of greater number of
tunnels (more porosity) which the $\L$CDM structure exhibits relative
to its $\T$CDM counterpart.  While the genus in case of $\T$CDM seems
extremely well-constrained (small scatter), the fluctuations in IMC,
at least at smaller \ffv$-$values ($<$0.15), are reasonably large.
This could be inferred as fluctuation in the sizes of the tunnels
forming due to the merger of the smaller clusters, while their total
number conserves for a given \ffv. ~The IMC and genus of the largest
cluster in the range \ffv$\in$[0.05,0.3] could be some of the best
discriminants of the two models.

\subsection{Morphology of the Superclusters}
This and the following subsection are devoted to a critical study of
the morphology of the ensemble of superclusters. Based on a
statistical study of many realizations of $\T$CDM, we are able to
establish it {\em for the first time} that the supercluster shapes and
their sizes are very well constrained measures, suitable as ideal
diagnostics of the large scale structure of the Universe and are
sensitive to the cosmological parameters of the model(s).

Since we are dealing with density fields, our superclusters are
defined as connected, overdense objects. In Section 4, we studied the
percolation properties of the density fields and found that the
percolation takes place at a threshold of density corresponding to
\dlt=1.7 in both $\L$CDM and $\T$CDM.  Further, \dlt$_{\rm max}\simeq$
3.5. Thus at the highest threshold of density, we are still within the
weakly nonlinear regime, and may find sufficiently large connected
objects which might act as dense progenitors of the percolating
superclusters. With this anticipation, we find it advisable to study
the morphology of the ensemble of superclusters as a function of the
density contrast over the entire range \dlt$\in$[0,\dlt$_{\rm max}$].
In so doing, we are also able to assess how the morphology of the LSS
changes with the density-level.

The morphology of the superclusters is quantified in terms of a set of
Shapefinders \citep{sss98,sheth03}. The dimensional Shapefinders have 
dimensions of length and are given by 
\b
\label{eq:tbl}
T = {3V\over S}; ~~~B = {S\over C}; ~~~L = {C\over4\pi(G+1)}.
\e
For objects with fairly simple topology, the above Shapefinders give
us a feel for the typical thickness (T), breadth (B) and length (L) of
the object being studied. Since we have demonstrated that most of the
superclusters are simply connected objects near or before the onset of
percolation, this interpretation is valid for structures defined in
this regime. Further, a combination of (T,B,L) can be used to define
Planarity (P) and filamentarity (F) of the object.
\b
\label{eq:pf}
P = {B-T\over B+T};~~~F={L-B\over L+B}.
\e

Individual superclusters in a given realization of, say, $\T$CDM
may follow a distribution of shapes and sizes, not necessarily
coinciding with a similar list due to another realization. Hence, to
make a statistical study of a given model, we average over the shapes
of individual superclusters defined at a given threshold of density,
and test whether the average morphology as a function of the
density-level is well constrained. For this purpose, we use the volume and
mass averaged Shapefinders defined as follows.
\begin{eqnarray}
\label{eq:avepf}
\bar{P}_{V,M}(\delta_{\rm th}) &=& {\sum_{i=1}^{N_C}{Q_i\times P_i}\over \sum_{i=1}^{N_C}Q_i}\nonumber\\
\bar{F}_{V,M}(\delta_{\rm th}) &=& {\sum_{i=1}^{N_C}{Q_i\times F_i}\over \sum_{i=1}^{N_C}Q_i},
\end{eqnarray}
where the quantity $Q_i$ may refer to the mass or volume of the
i$^{\rm th}$ supercluster defined at the level \dlt$_{\rm th}$, and the
summation is over all the objects. It is to be noted here that the
average planarity and filamentarity are mostly contributed by the
structures with large volume and/or mass. There are a finite number of
such objects before the onset of percolation. Hence, in this regime,
we probe the {\em average shape} of these objects. However, once the
percolation sets in, the average morphology mainly reflects the
morphology of the largest object in the system.

We evaluate $\bar{P}_{V,M}$ and $\bar{F}_{V,M}$ for all the 11 samples at the
density levels corresponding to \dlt$\in$[0,\dlt$_{\rm max}$]. The
measurements due to 10 $\T$CDM realizations are used for averaging and
for evaluating the errors.
\begin{figure*}
  \centering
  \centerline{
    \includegraphics[scale=0.8]{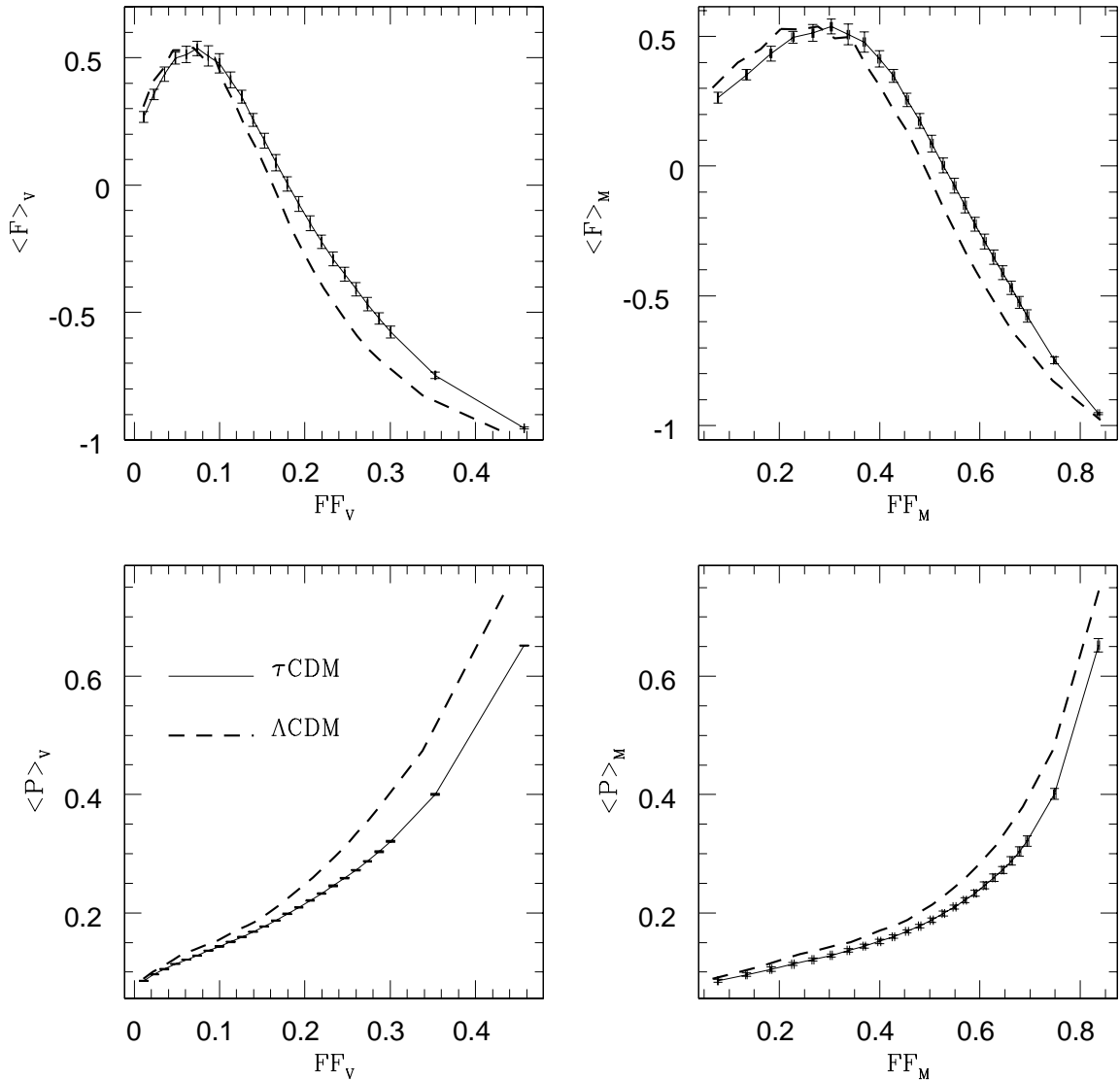}
    }
  \caption{The dimensionless Shapefinders (P,F) are evaluated for all the clusters
    at a given density level \dlt$_{\rm th}$. These Shapefinders are
    volume-averaged and are studied w.r.t. \ffv~(left panels). They
    are also mass-averaged and are studied w.r.t. \ffm~(right panels).
    The average shape of the cosmic web assumes increasingly planar
    morphology as \dlt$_{\rm th}$ is lowered. The corresponding
    negative filamentarity is due to the presence of large voids which
    lead to the development of appreciable negative curvature in the
    system. The average filamentarity in the system is highest near
    the onset of percolation, thus making it ideal for the study of
    structures which are evident in the visual impression of the
    cosmic web. The average morphology revealed in terms of
    $\bar{P}_{V,M}$ and $\bar{F}_{V,M}$ is extremely well constrained for a
    given model and could be a useful diagnostic for discriminating
    the models of structure formation.}
\label{fig:meanPF}
\end{figure*}
Figure \ref{fig:meanPF} shows $\bar{P}_{V,M}$ and $\bar{F}_{V,M}$ studied as
functions of \ffv ~(left panels) and \ffm ~(right panels)
respectively. The solid lines refer to $\T$CDM and the dashed lines
refer to $\L$CDM. The vertical bars stand for the 1$\sigma$ errors. We
observe that
\begin{itemize}
\item $\bar{P}_{V,M}$ and $\bar{F}_{V,M}$ are well constrained statistics and
  hence, can be used to reliably assess the morphology of the LSS.
\item The average planarity of the LSS steadily increases as we lower
  the threshold of density.
\item The average filamentarity initially increases, reaches a maximum
  where the percolation sets in, but drops to sub-zero values after
  \ffv$\sim$0.18.
\end{itemize}
This implies that the square of the area of the excursion set is {\em
  always} greater than thrice the product of the volume and curvature.
Further, C$^2$ is superseded by 4$\pi$(G+1)S at \ffv$>$0.18. A steady
drop in the curvature is responsible for this behaviour. We note that
the system also tends to become increasingly spongy. Taken together,
these observations lead to the following picture: as the density level
is brought down, the overdense structures connect up across the
sample-volume. This at a time leads to large scale filaments and also
to the birth of large voids in the system, signifying the onset of
percolation. As revealed in the top panels, the filamentarity of the
system is highest at the time. With a further drop in \dlt, the
filaments surrounding the ``voids'' grow thicker and locally these
assume a slab-like structure. Such an excursion set, at a time has
large planarity, but also has large negative curvature due to being
surrounded by large, almost spherical voids. This negative
curvature is responsible for the large negative filamentarity
{\footnote{It was anticipated that the Shapefinders (P,F)$\in$[0,1].
    Our present exercise shows that excursion sets corresponding to
    \dlt$<$1 show negative filamentarity. These objects are
    necessarily dominated by large negative curvature.}}.

We notice that both the statistics are extremely well-constrained.
Further, $\bar{P}_{V,M}^{\L} \ge \bar{P}_{V,M}^{\T}$ and
$\bar{F}_{V,M}^{\T} \ge \bar{F}_{V,M}^{\L}$. This feature can be used
to distinguish these two models, and is an instance of the fact that
morphology of LSS is an equally robust statistic as MFs to compare LSS
due to rival models of structure formation.

We finally note that the superclusters defined according to the
conventional definition (\dlt$_{SC}\in$[1,2]) do correspond to the
percolating structures in the system, and indeed exhibit the highest
filamentarity (the top panels of Figure \ref{fig:meanPF}).  This
proves that the Shapefinder statistics conform to our visual
impression while quantifying the LSS. In the next subsection, we shall
try to quantitatively asses the sizes of these superclusters and shall
demonstrate that these measures are sensitive to the models being
investigated, enabling us to distinguish the two models successfully.

\subsection{Cumulative Probability Functions}
In this subsection we make a quantitative investigation of the sizes
of the superclusters and study the variation of these measures as
functions of the density contrast. Our main focus is to make a
statistical study of these measures and to develop a methodology to
handle various statistics due to many realizations of a given model.
We also compare the sizes of superclusters in $\T$CDM with those in
$\L$CDM and find that these are sensitive to the model being
investigated and can be used to distinguish the two models.

For reasons explained earlier, we find it advisable to carry out the
exercise over a set of levels of density. We study the sizes of
superclusters starting from the highest density level
(\dlt$\simeq$3.5) down to the percolation threshold.

Since the sizes of individual structures may vary from one realization
to another, it is not meaningful to prepare a list of, say, top N
superclusters in a given realization and check its statistical
significance over other realizations.  There may not be any unique
correspondence of the rank of the supercluster in the list and its
size. This is mainly because at the density-levels above or near the
percolation threshold, there may be many structures with comparable
sizes and masses. So the sorted array of structures in this range of
density-levels will show a large amount of Poisson fluctuations in the
size of, say, the i$^{\rm th}$ largest cluster. Hence, a more
meaningful approach is to study the cumulative probability functions
of the Shapefinders so as to know how many superclusters in a given
ensemble of them, have their sizes greater than a given triplet
(T,B,L).

As noted in Section 4, all the 10 samples of $\T$CDM being studied
here have comparable number of clusters at any given threshold of
density, and that these are only marginally smaller in number than in
$\L$CDM.  Further, we found that the global geometry and topology of
the $\T$CDM density fields is extremely well constrained. Hence, it is
justified to construct cumulative probability functions (hereafter,
CPFs) of various quantities using an ensemble which contains the
clusters due to {\em all} the 10 realizations of $\T$CDM
{\footnote{However, prior to this, we have checked that the CPFs due
    to individual realizations of $\T$CDM show trends similar to CPFs
    due to the cumulative ensemble.}}. The CPF of a given quantity Q
is a fraction of the total number of structures having the value of Q
greater than a chosen value q. In our case, the total number of
clusters amounts to the sum over all the realizations of $\T$CDM and
is close to a few thousand.  It turns out that it is beneficial to
work with such large number of clusters, so that we have sufficient
number of structures which are otherwise rare in a given sample. By
finding these in sufficient number in a bigger ensemble, we can put an
upper limit on the sizes of superclusters at any given threshold of
density, without hampering our conclusions by Poisson noise.

To begin with, we study the CPFs of the dimensional Shapefinders, the
so called, Thickness (T), Breadth (B) and Length (L) of the
superclusters with \ffv ~as a parameter.
\begin{figure*}
  \centerline
  \centering{
    \includegraphics[scale=0.7]{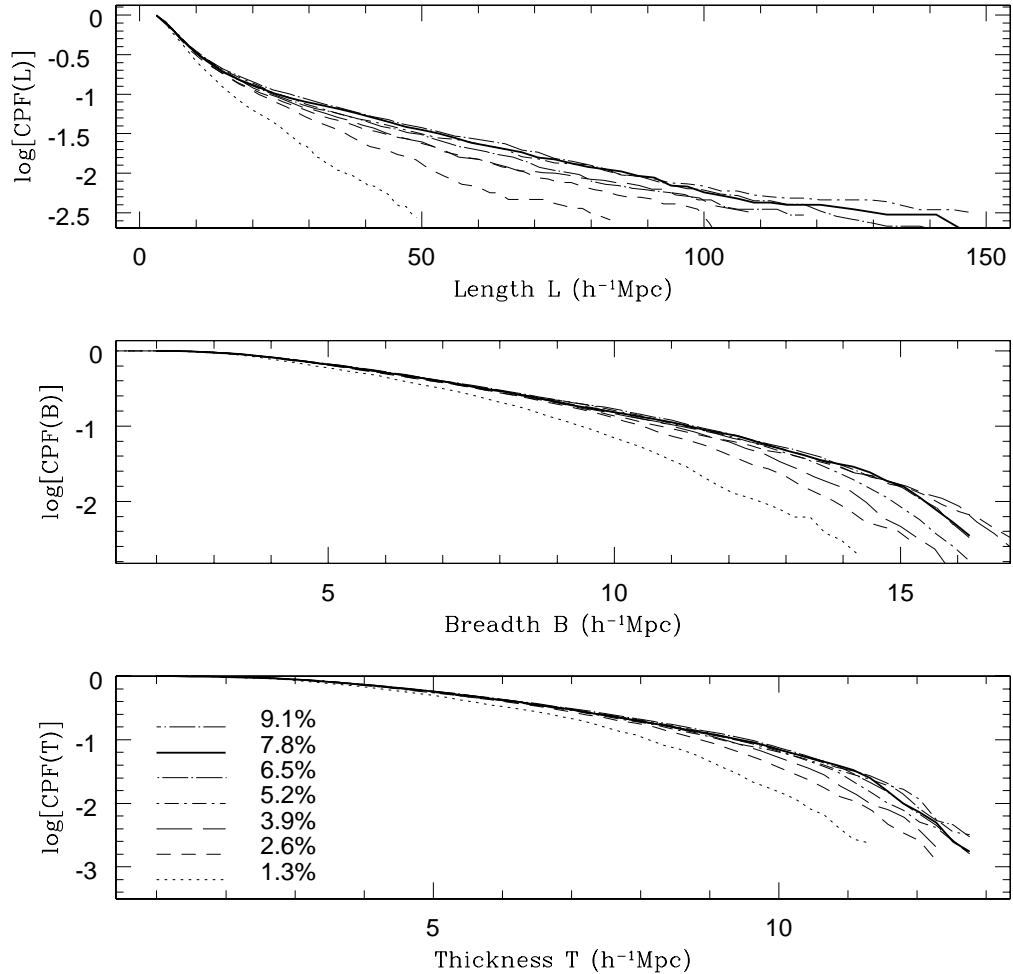}
    }
  \caption{The CPFs of dimensional Shapefinders are studied as functions of
    the density level for $\T$CDM. The bottom panel shows the
    associated volume filling fractions. The percolation curves are
    shown as bold, solid lines. We notice that the clusters grow
    thicker, broader and longer as we increase the volume fraction (at
    any given triplet (T,B,L), there are more number of them thicker
    than T, broader than B and longer than L as we decrease \dlt).}
  \label{fig:evolcpf}
\end{figure*}
Figure \ref{fig:evolcpf} shows the results. We notice that the CPFs
follow a specific pattern of evolution. While a typical range of
thickness and breadth of the structures is fixed (T $\in$(0,13)
h$^{-1}$Mpc and B $\in$ (0,17) h$^{-1}$Mpc), the fraction of clusters
with a given thickness and breadth increases with increasing \ffv.
Thus, the structures become thicker and broader as we lower the
threshold, as anticipated. The CPF of length, on the other hand,
extends to high L$-$values upon increasing \ffv. This implies that the
longer structures become statistically more significant with the
decrease of \dlt, while their thickness and breadth increases only
marginally. Thus, as \dlt$\rightarrow$\dlt$_{\rm perc}$, the
filamentarity of the individual superclusters should increase. At the
highest thresholds, however, the superclusters might look only
marginally filamentary (F$\sim$0.25, P$\sim$0.1) (see, Figure
\ref{fig:meanPF}), and in realistic situation, may give us an
impression of being more like {\em ribbons}. {\footnote{A more
    detailed study of the morphology in this range of \dlt ~needs to
    be carried out to understand whether the super-structures with
    significant planarity are statistically significant or not. This
    question has an observational bearing because some of the
    super-structures in our local neighbourhood appear planar. For
    example, the Great Wall, is characterised by (T,B,L)$\simeq$(10,
    50, 120) h$^{-1}$Mpc.  See, Fairall 1998 and Martinez \& Saar
    2002.}}. We infer that the top 10 superclusters in $\T$CDM should
have their length $\ge$40 h$^{-1}$Mpc at \dlt$\sim$\dlt$_{\rm max}$,
and these should grow to become as long as L$\ge$90h$^{-1}$Mpc at the
onset of percolation. The longest superclusters at percolation, which
are necessarily rare (say, a few in 1000), might be as long as
150h$^{-1}$Mpc.

To summarise, the supercluster-sizes are dependent on the threshold of
density at which the superclusters are defined. Based on the study of
CPFs as functions of \dlt$_{\rm th}$ we infer that the thickness and
breadth of the structures are comparable to one another at all
thresholds of density \dlt$\le$\dlt$_{\rm perc}$, with breadth B {\em
  always} slightly higher than thickness T. We have noted that at
percolation, the largest superclusters could be as long as 150
h$^{-1}$Mpc, and the superclusters with length L$\in$[90,150]
h$^{-1}$Mpc are relatively more frequently to be found (say, with a
few per cent probability in an ensemble of 1000 structures).

We are limited by the availability of only one realization of $\L$CDM.
Hence similar conclusions cannot be reliably drawn for this model, but
a study of CPFs of $\L$CDM superclusters does reveal similar trends.

In order to see whether the supercluster sizes and their morphology
can be utilised to distinguish the two models, we studied the CPFs of
T, B and L due to $\L$CDM and $\T$CDM together.
\begin{figure*}
  \centerline \centering{ \includegraphics[scale=0.7]{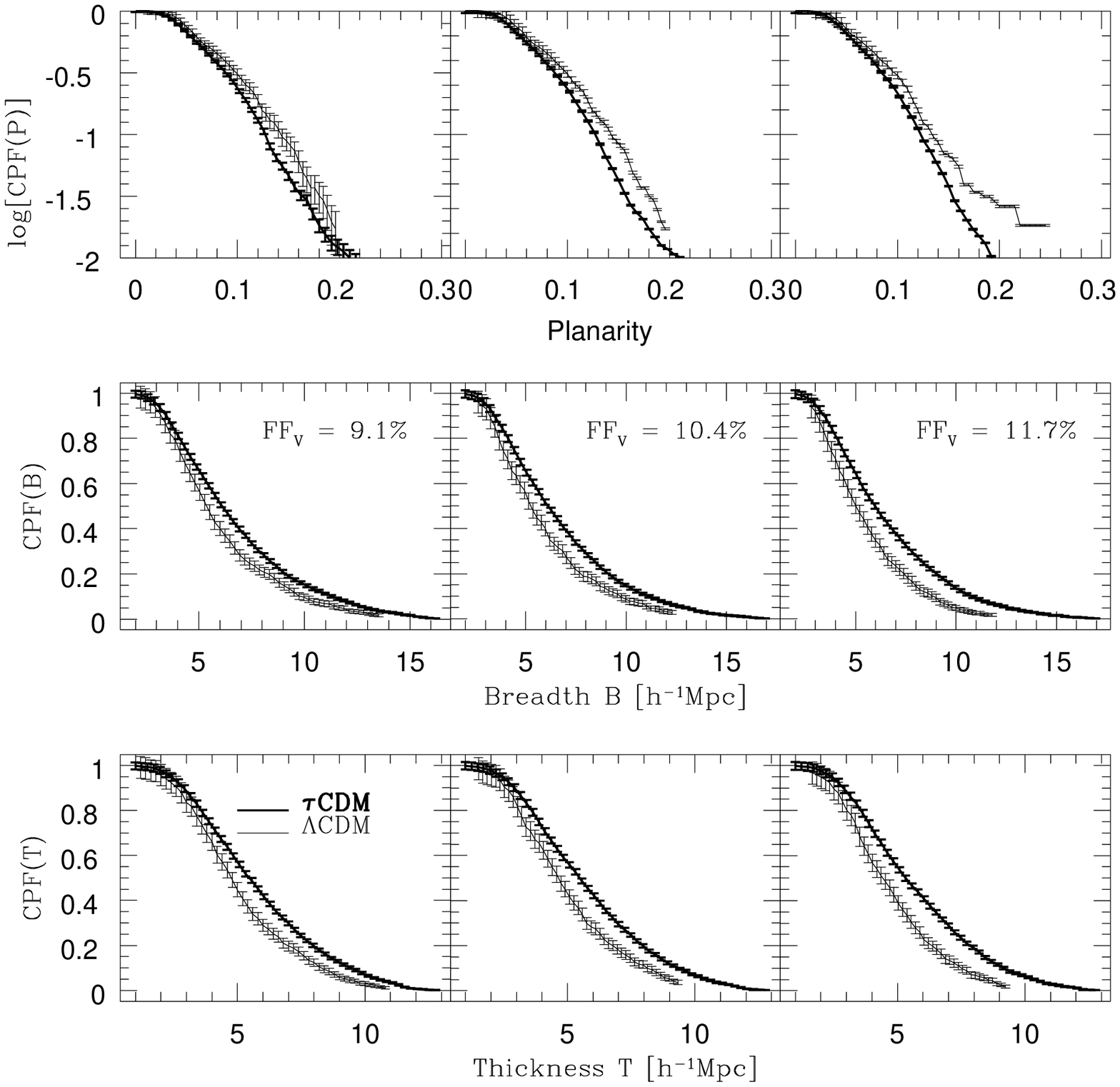} }
  \caption{The cumulative probability functions for thickness (T), 
    breadth (B) and planarity (P) are shown for both the models at
    three thresholds of density.  The fraction of clusters with a
    given thickness or breadth in $\T$CDM are {\em always} greater
    than or equal to the fraction of clusters in $\L$CDM. Further
    CPF$^{\L}$(P)$\ge$CPF$^{\T}$(P).  These Shapefinders can hence, be
    used to distinguish these models.}
\label{fig:tbcpf}
\end{figure*}
We find that near to the onset of percolation, the two dimensional
Shapefinders thickness and breadth, and the planarity of the
superclusters do give a detectable signal. Figure \ref{fig:tbcpf}
shows the results. The thick solid lines refer to $\T$CDM. The
1$\sigma$ error-bars are computed assuming a Poisson statistics and go
as the square-root of the number of clusters available at a given
value of (T,B,P).  It has been checked that the CPFs due to individual
realizations fall within 1$\sigma$ of the CPFs due to the total
ensemble in most cases.  Thus our decision to use the total ensemble
(cumulative of all the clusters and due to all the realizations) and
of assuming Poisson errors is justified. To make the comparison
ideal, we choose to incorporate Poisson errors on the CPFs of the only
available realization of $\L$CDM as well. As shown in the figure, the
$\T$CDM CPFs of thickness and breadth are systematically higher than
those of $\L$CDM at \ffv=9.1, 10.4 and 11.7 per cent. Thus, we are
able to establish that the $\T$CDM superclusters with a given
thickness are statistically larger in number than in $\L$CDM. Assuming
similar number of clusters in the two models, we could infer that the
$\T$CDM superclusters are thicker than those in $\L$CDM.  This effect
is pronounced enough to be detected with a set of limited number of
realizations. The topmost panels further reveal that
CPF$^{\T}$(P)$\le$CPF$^{\L}$(P). We conclude from here that the
thickness, breadth and planarity of superclusters follow well-defined
distributions which are special to the model being studied.

\begin{figure*}
  \centerline
  \centering
  {\includegraphics[scale=0.9,trim=20 140 5 140,clip]{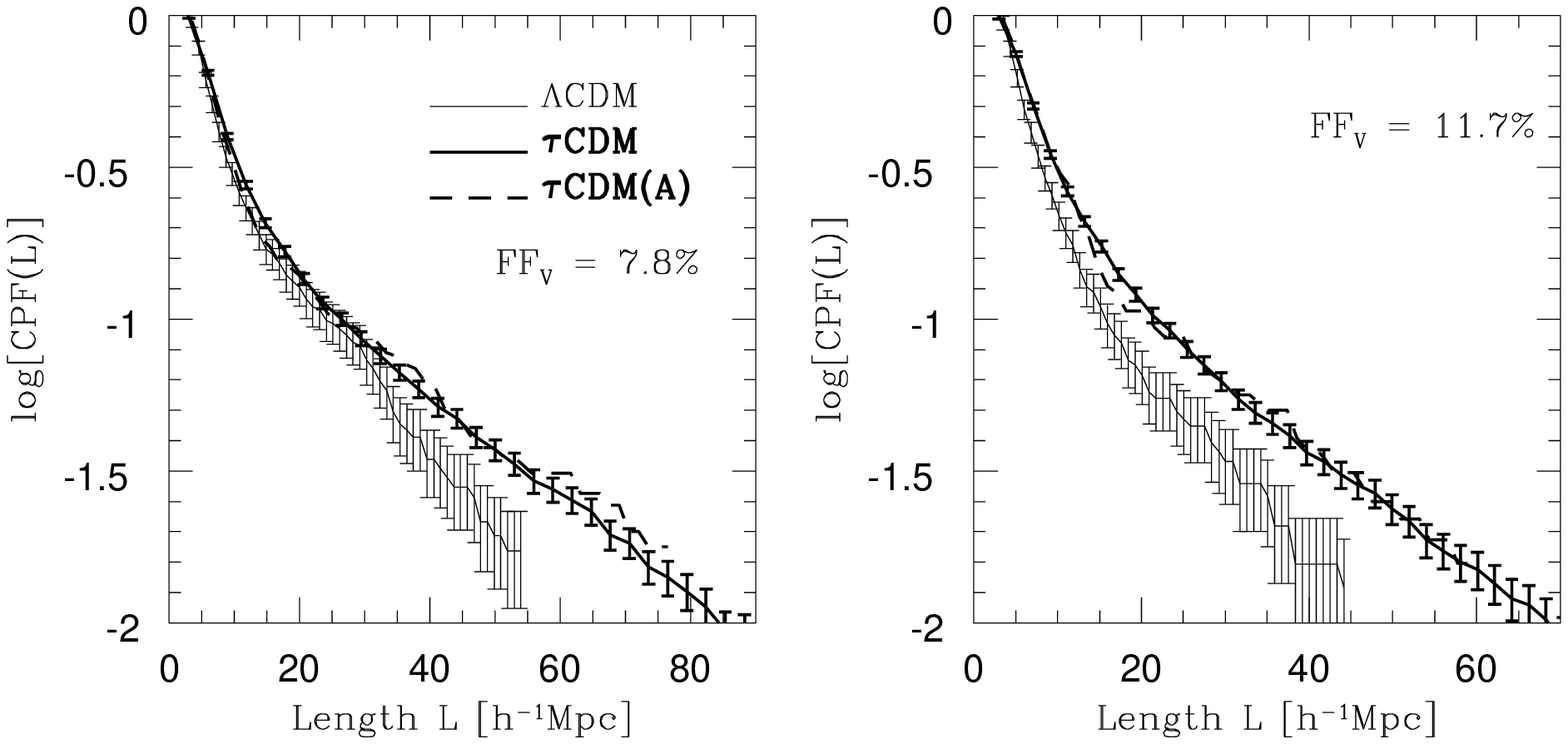}
    }
  \caption{{Shown here are the cumulative probability functions of
      length L for clusters of $\T$CDM and $\L$CDM at two thresholds
      of density corresponding to \ffv=7.8 per cent (just before the
      onset of percolation) and \ffv=11.4 per cent (after the onset of
      percolation). We find the CPFs of the two models to be
      distinctly different at longer length-scales. The dashed line
      refers to the CPFs due to the first realization of $\T$CDM which
      shares the initial set of random numbers with $\L$CDM. The
      largest superclusters of $\T$CDM are systematically larger than
      those due to $\L$CDM.}}
  \label{fig:cpflen}
\end{figure*}
In order to test whether we can constrain the length of superclusters
of a given model, we studied the CPF for length L, following identical
procedure. Figure \ref{fig:cpflen} shows the results, where the CPF(L)
have been evaluated at \ffv=7.8 per cent (at the onset of percolation)
and at \ffv=11.7 per cent (after the percolation). We notice that at
the onset of percolation, the length of the longest $\T$CDM
superclusters could be as large as 90 h$^{-1}$Mpc, whereas their
$\L$CDM counterpart structures, which exhibit same degree of
statistical significance, are relatively shorter with L$_{\rm max}\ge$
55 h$^{-1}$Mpc. We conclude that the $\T$CDM superclusters tend to be
statistically longer than their $\L$CDM counterpart structures. Thus,
we can also utilise the length of the superclusters to compare and
distinguish the two models. After the percolation, the longest
structure dominates the survey volume and enters only as a Poisson
fluctuation in the CPF of length. The remaining structures are
evidently shorter in length compared to their longer progenitors at
percolation, so that CPF(L) are confined to lower values of L. This
effect is evident in the right panel of Figure \ref{fig:cpflen}. In
both the panels, shown as dashed line is the CPF(L) due to the first
realization of $\T$CDM which shares its set of initial random numbers
with the $\L$CDM simulation. We note the sharp tendency of $\T$CDM
structures to be longer than those of $\L$CDM, an effect which
successfully enables us to distinguish the two models of structure
formation.
\section{Discussion and Conclusions}
This paper provides a theoretical framework and a methodology to
analyse the galaxy-distribution revealed by large 3-dimensional
redshift surveys. The exercise is in the wake of preparing to tackle
the redshift surveys like 2dFGRS and SDSS, and consists in studying
the geometry, topology and morphology of the LSS as revealed by the
volume limited samples derived from 10 realizations of mock
$\T$CDM$-$based SDSS catalogue. These volume limited samples are
prepared so as to refer to the same physical volume, and exhibit
similar number density of galaxies. The density fields used in our
calculations are prepared after smoothing all the samples at 6
h$^{-1}$Mpc.

We use SURFGEN (Sheth et al.2003) for calculating MFs, and find MFs as
well as the derived morphological statistics, the Shapefinders, to be
well-constrained statistics, useful to probe the contribution of the
higher order correlation functions on the clustering of the
galaxy-distribution. For this purpose, we compare the measured MFs due
to 10 realizations of $\T$CDM with those due to a realization of
$\L$CDM.  All the 11 realizations, by construction, reproduce the
observed two$-$point correlation function measured using the APM
catalogue, and show similar clustering amplitude.  Some of the
important results of our analysis are summarised below:
\begin{itemize}
\item We show that the global MFs (defined as the sum of partial MFs
  over the set of all the clusters at a given threshold of density) of
  $\T$CDM show systematically lower amplitude compared to those due to
  $\L$CDM, an effect which enables us to distinguish the two models,
  and can be attributed to the nonzero higher order correlation
  functions. Considering the fact that higher order correlation
  functions are cumbersome to estimate and offer little insight in
  their interpretation, this can be termed as a success of MFs as
  efficient quantifying statistics of LSS.
\item The Virgo simulations of dark matter showed {\em higher}
  amplitude of MFs for $\T$CDM compared to $\L$CDM (see Sheth et
  al.2003). The cosmological parameters used by Virgo group match with
  those employed by CHWF except for the shape parameter \Gm, which is
  0.21 in the former case and 0.25 in the latter. We assume that this
  minute difference in \Gm ~will not alter the above trend so long as
  both the models start with the same power spectrum. The {\em galaxy
    distributions} due to both the models were analysed in {\it this} paper,
  and we found a reversal in the above trend: The MF-amplitudes for
  $\T$CDM were found to be {\em smaller} than $\L$CDM.
  
  We note that the relative smallness of the amplitudes of MFs
  reflects the higher degree of phase correlations in the
  matter-distribution.  Taken together with this fact, our above
  observation points out that scale-dependent biasing could lead to an
  apparent phase-mismatch between the galaxy-distribution and the
  underlying matter-distribution. Thus, there may be {\em larger} degree of
  phase-correlations in the {\it matter-distribution} of $\L$CDM
  compared to $\T$CDM, but the biased {\it galaxy-distribution} of
  $\L$CDM would appear to exhibit {\em smaller} phase-correlations compared
  to $\T$CDM. We note however, that these results are affected by the
  constraint that the bias is fixed so as to reproduce the observed
  2$-$point correlation function (see Section 2).
  
  An analysis of 2dFGRS performed by \cite{lahav2df02} concluded that
  the average bias on scales 0.02 $\le k \le$ 0.15 h Mpc$^{-1}$ is
  indeed $\sim$1 {\footnote{However, it should be noted that the
      cosmological model was assumed to be $\L$CDM.}}. In light of the
  above fact, it will be interesting to see how the geometry, topology
  and morphology of the LSS change in the presence of
  scale-independent, constant bias.  From our analysis, however, it is
  clear that bias in various structure formation scenarios can play
  important role, and comparing the {\em dark matter distributions}
  with the SDSS and/or 2dFGRS galaxy distributions {\em is not} an
  advisable way of testing the model(s) of structure formation against
  the observations.  A realistic treatment of bias {\it en route} the
  proper treatment of the physics of galaxy formation in preparing a
  mock catalogue, is desired in any such exercise.
  
\item Since we had access to 10 realizations of $\T$CDM, we made a
  statistical study of the shapes and sizes of the superclusters
  occurring in this model. We found the thickness and breadth of the
  superclusters to increase only marginally over a range of \dlt ~(T,
  B$\in$[1,17]h$^{-1}$Mpc; T$\le$B). However, the length of the
  largest superclusters increases monotonically from $\sim$40
  h$^{-1}$Mpc to $\sim$90 h$^{-1}$Mpc as \dlt$\rightarrow$\dlt$_{\rm
    perc}$. Near the percolation, the $\T$CDM superclusters with
  length $\ge$90 h$^{-1}$Mpc are $\sim$ 1 per cent probable to be
  found (say, the top 10 in an ensemble of 1000 structures).  However,
  the longest (and necessarily the most massive) superclusters could
  be as long as 150 h$^{-1}$Mpc and are rare to be found, with less
  than 1 per cent probability in a volume as big as that covered in
  this analysis. These findings are to be confronted with recent
  observational work by \cite{brand03} who report rare structures of
  the size of $\sim$100 Mpc being traced by radio galaxies. The
  dominant morphology of the superclusters is prolate, or ribbon-like
  at the highest threshold of density, and it evolves to become more
  filamentary as the threshold is lowered toward the percolation
  threshold.
\item We further found that the CPFs of thickness and breadth near the
  percolation threshold show distinctly higher amplitude in $\T$CDM
  compared to $\L$CDM. Even planarity of the two models shows
  appreciably different distribution for the two models. These
  Shapefinders, thus, could be good discriminants of the structure
  formation scenarios.
\item A study similar to the above, repeated for the length of the
  superclusters reveals that the longest superclusters of $\T$CDM are
  likely to be longer than their $\L$CDM counterpart structures at the
  same level of significance. In particular, whereas the longest
  superclusters of $\T$CDM could be as long as 90 h$^{-1}$Mpc, the
  $\L$CDM superclusters have L$\ge$55 h$^{-1}$Mpc.
\end{itemize}

To summarise, we have established that the Minkowski Functionals
estimated using SURFGEN are well constrained statistics and can be
reliably used to quantify the geometry and topology of the LSS in
various scenarios of structure formation.  We have used these to
distinguish two rival models of structure formation, namely $\L$CDM
and $\T$CDM, and have explored a set of related measures and
morphological statistics like Shapefinders to make such a comparison
more robust. We have further measured and reported the sizes and
shapes of the superclusters in $\T$CDM, with a scope of extending
similar methodology to other models with the availability of larger
number of realizations for them.

Our present exercise does have its limitations, which are mentioned
below. 
\begin{itemize}
\item The effect of redshift space distortions has not been
  incorporated in the present exercise. We have throughout worked in
  the redshift space and have not quantified the {\em fingers-of-God}
  effect. Such a treatment is beyond the scope of this paper and shall
  be taken up in a future work.
\item The mock catalogues studied in this paper are prepared by
  implementing {\em ad-hoc} biasing schemes. Although the methodology
  of analysis is clearly laid out in the present work, a more
  realistic comparison of the mock data with the observed LSS of the
  Universe should be carried out using the mock catalogues prepared by
  incorporating the physics of galaxy-formation in as much detail as
  possible. The semi-analytic methods, the SPH simulations of galaxy
  formation and the Conditional Luminosity Function based approach
  should be able to deliver such mock catalogues in near future.
\end{itemize}
We hope to address some of the issues mentioned above and extend this
work to a comparison of the mock catalogues with the LSS revealed by
2dFGRS and SDSS in near future.
\section{Acknowledgements}
JVS wishes to thank Varun Sahni for his constant support,
encouragement and for his critical comments on the manuscript. Thanks
are due to Sergei Shandarin for his encouragement and for the numerous
discussions with him from which the author has benifited a lot. It is
a pleasure to recall stimulating discussions with Somnath Bharadwaj,
Ranjeev Misra and K.Subramanian.  JVS also wishes to thank the CHWF
team for making the mock data public, and in particular, Shaun Cole
who has, time and again, provided useful information pertaining to
data in the course of this work. Thanks are due to Matteo Frigo and
Steven Johnson for making FFTW public.  JVS is supported by the Senior
Research Fellowship of the Council of Scientific and Industrial
Research (CSIR), India. The data analysed in this paper are available
at {\it http://star-www.dur.ac.uk/$\sim$cole/mocks/main.html} .


\begin{thebibliography}{99}
  
\bibitem[\protect\citeauthoryear {{Abazajian et al.}}{2003}]{abaz03}
  {Abazajian}, K., {Adelman-McCarthy}, J.K., {Agueros}, M.A., {Allam},
  S.S. \& the SDSS Collaboration, 2003, astro-ph/0305492

\bibitem[\protect\citeauthoryear{{Basilakos}}{2003}]{basil03}{Basilakos},S.
  2003, astro-ph/0302596, Accepted for publication in MNRAS

\bibitem[\protect\citeauthoryear{{Baugh}}{1996}]{baugh96} {Baugh},C.M.,
  1996, MNRAS, 280, 267
  
\bibitem[\protect\citeauthoryear{{Baugh} \&
    {Efstathiou}}{1993}]{bauef93} {Baugh},C.M., {Efstathiou}, G., 1993,
  MNRAS, 270, 183

\bibitem[\protect\citeauthoryear{{Bennett et al.}}{2003}]{bennett03}
  {Bennett}, C.L., et al., 2003, ApJS, 148, 1
  
\bibitem[\protect\citeauthoryear{{Benson et al.}}{2002}]{benson02}
  {Benson}, A.J., {Lacey}, C.G., {Baugh}, C.M., {Cole}, S., {Frenk},
  C.S., 2002a, MNRAS, 333, 156

\bibitem[\protect\citeauthoryear{{Berlind et al.}}{2003}]{berlind02}
  {Berlind},A., {Weinberg}, D.H., {Benson}, A.J., et al., astro-ph/0212357

\bibitem[\protect\citeauthoryear{Bharadwaj, et. al. }{2000}]{bharad}
  {Bharadwaj}, S., {Sahni}, V., {Sathyaprakash}, B.S., {Shandarin},
  S.F., and {Yess}, C., 2000, ApJ, 528, 21.

 \bibitem[\protect\citeauthoryear{Blanton et. al.}{2003}]{blant03}
  {Blanton}, M., {Hogg}, D., {Bahcall}, N., 2003, ApJ, 592, 819

\bibitem[\protect\citeauthoryear{{Bond} \& {Efstathiou}}{1991}]{be91}
  {Bond}, J.R. and {Efstathiou}, G., 1991, Phys.Lett.B, 265, 245

\bibitem[\protect\citeauthoryear{{Brainerd \& Specian}}{2003}]{brainerd03}
  {Brainerd}, T. \& {Specian}, M., 2003, ApJL, 593, 7

\bibitem[\protect\citeauthoryear{{Brand et al.}}{2003}]{brand03}
  {Brand}, K., {Rawlings}, S., {Hill}, G., {Lacy}, M., {Mitchell}, E.,
  \& {Tufts}, J., 2003, MNRAS, 344, 283

\bibitem[\protect\citeauthoryear{{Couchman}}{1991}]{couch91}{Couchman},
  H. M. P., 1991, ApJ, 368, 23
  
\bibitem[\protect\citeauthoryear{{Cole}, {Hatton}, {Weinberg}, \&
    {Frenk}}{1998}]{cole2df98} {Cole}, S., {Hatton}, S., {Weinberg},
  D.H., {Frenk}, C.S., 1998, MNRAS, 945
  
\bibitem[\protect\citeauthoryear{{Cole et al.}}{2000}]{cole00} {Cole},
  S., {Lacey}, C.G., {Baugh}, C.M., {Frenk}, C.S., 2000, MNRAS, 319,
  168
  
\bibitem[\protect\citeauthoryear{{Colless et al.}}{2003}]{colless03}
  {Colless}, M. and the 2dFGRS team, astro-ph/0306581

\bibitem[\protect\citeauthoryear{{Colley} et al.}{2000}]{colley00}{Colley}, W.N., {Gott}, J. R. III, {Weinberg}, D. H.,
  {Park},C. and {Berlind}, A. A., 2000, ApJ, 529, 795

\bibitem[\protect\citeauthoryear{Doroshkevich et al.}{2003}]{dorosh03}
  {Doroshkevich}, A., {Tucker}, D.L., {Allam}, S. \& {Way}, M.J., astro-ph/0307233,
  Submitted to Astronomy \& Astrophysics
  
\bibitem[\protect\citeauthoryear{{Einasto et al.}}{2003}]{einasto03}
  {Einasto}, J., {H''utsi}, G., {Einasto}, M. et al., 2003, Astronomy \&
  Astrophysics, 405, 425
  
\bibitem[\protect\citeauthoryear{{Fairall}}{1998}]{fairall}
  {Fairall}, A., 1998, {\it Large-Scale Structure in the Universe}, John-Wiley \&
  Sons in association with Praxis Publishing, Chichester
  
\bibitem[\protect\citeauthoryear{{Gott}, {Melott} \&
    {Dickinson}}{1986}] {gott86} {Gott}, J.R., {Melott}, A.L. \&
  {Dickinson}, M., 1986, ApJ, 306, 341

\bibitem[\protect\citeauthoryear{{Gott et al.}}{1989}]{gott89}
  {Gott}, J.R., {Miller}, J., {Thuan}, T. X., et al., 1989, ApJ, 
  340, 625

\bibitem[\protect\citeauthoryear{{Hamilton et al.}}{1991}]{hamilton91}
  {Hamilton}, A.J.S., {Kumar}, P., {Lu}, E., {Matthews}, A., 1991, ApJL, 
  374, 1
  
\bibitem[\protect\citeauthoryear{{Hikage} et
    al.}{2003}]{hikage03a}{Hikage},C.  {Schmalzing}, J., {Buchert},
  T., {Suto}, Y., {Kayo}, I., {Taruya},A., {Vogeley}, M., {Hoyle}, F.,
  {Gott}, J. R. III, {Brinkmann}, J. et al., astro-ph/0304455,
  Accepted for publication in PASJ
  
\bibitem[\protect\citeauthoryear{{Hikage}, {Taruya} \& {Suto}}{2003}]{hikage03b}
  {Hikage}, C., {Taruya}, A.\& {Suto}, Y., 2003, PASJ, 55, 335

\bibitem[\protect\citeauthoryear{{Jenkins} et al.}{1998}]{jenkcdm}
  {Jenkins}, A.R. et al. (for the Virgo Constortium), 1998, ApJ, 499,
  20 
  
\bibitem[\protect\citeauthoryear{Klypin \& Shandarin}{1993}]{ksh93}
  Klypin, A.A. \& Shandarin, S.F., 1993, ApJ, 413, 48 
  
\bibitem[\protect\citeauthoryear{{Kneib et al.}}{2003}]{kneib03}
  {Kneib}, J-P., {Hudelot}, P., {Ellis}, R., et al., astro-ph/0307299,
  accepted for publication in ApJ

\bibitem[\protect\citeauthoryear{{Kravtsov et al.}}{2003}]{kravtsov03}
  {Kravtsov}, A.V., {Berlind}, A., {Wechsler}, R. et al., 2003,
  astro-ph/0308519

\bibitem[\protect\citeauthoryear{{Lahav et al.}}{2002}]{lahav2df02}
  {Lahav}, O., {Bridle}, S., {Percival}, W.J., et al. and the 2dFGRS team,
  2002, MNRAS, 333, 961
  
\bibitem[\protect\citeauthoryear{{Lahav} \&
    {Suto}}{2003}]{lahav-suto03} {Lahav}, O. \& {Suto}, Y.,
  astro-ph/0310642, invited review article submitted to Living Reviews
  in Relativity

\bibitem[\protect\citeauthoryear{{Magliocchetti} \& {Porciani}}{2003}]{maglio03}
  {Magliocchetti}, M. \& {Porciani}, C., 2003, astro-ph/0304003, Accepted for 
  publication in MNRAS
  
\bibitem[\protect\citeauthoryear{{Martinez}\&{Saar}}{2002}]{martinez02}
  {Martinez}, V. \& {Saar}, E., 2002, Chapman \& Hall/CRC
    
\bibitem[\protect\citeauthoryear{{Matsubara}}{2003}]{matsubara03}
  {Matsubara}, T., 2003, ApJ, 584, 1

\bibitem[\protect\citeauthoryear{{Mecke}, {Buchert} \&
    {Wagner}}{1994}]{meckwag94} {Mecke}, K.R., {Buchert}, T. \&
  {Wagner}, H., 1994, A$\&$A, 288, 697
  
\bibitem[\protect\citeauthoryear{{Melott}}{1990}]{melotopo90}
  {Melott}, A.L., 1990, Phys. Rep., 193, 1

\bibitem[\protect\citeauthoryear{{Melott} \& {Dominik}}{1993}]{melot93}
  {Melott}, A.L. \& {Dominik}, K., ApJL, 86, 1

\bibitem[\protect\citeauthoryear{{Norberg et al.}}{2002a}] {norberg02a}
  {Norberg}, P.,{Cole}, S., {Baugh}, C.M. and the 2dFGRS team, 2002a, MNRAS, 336,
  907

\bibitem[\protect\citeauthoryear{{Norberg et al.}}{2002b}] {norberg02b}
  {Norberg}, P., {Baugh}, C.M., {Hawkins}, E., and the 2dFGRS team, 2002b, MNRAS,
  332, 827

\bibitem[\protect\citeauthoryear{{Peacock} \& {Dodds}}{1994}]{peadod94}
  {Peacock}, J. A. and {Dodds}, S. J., 1994, MNRAS, 267, 1020

\bibitem[\protect\citeauthoryear{{Sahni} \& {Coles,}}{1995}]{sc95}
  {Sahni}, V. and {Coles}, P., 1995, Phys.Rep., 262, 1
  
\bibitem[\protect\citeauthoryear{Sahni, Sathyaprakash \&
    Shandarin}{1997}]{sss97} {Sahni}, V., {Sathyaprakash}, B.S. \&
  {Shandarin}, S.F., 1997, ApJL, 476, L1
  
\bibitem[\protect\citeauthoryear{Sahni, Sathyaprakash \&
    Shandarin}{1998}]{sss98} {Sahni}, V., {Sathyaprakash}, B.S. \&
  {Shandarin}, S.F., 1998, ApJL, 495, L5
  
\bibitem[\protect\citeauthoryear{Sathyaprakash, Sahni \&
    Shandarin}{1998}]{sss98b} {Sathyaprakash}, B.S., {Sahni}, V. \&
  {Shandarin}, S.F., 1998, ApJ, 508, 551
  
\bibitem[\protect\citeauthoryear{Sathyaprakash, Sahni \&
    Shandarin}{1996}]{sss96} {Sathyaprakash}, B.S., {Sahni}, V. \&
  {Shandarin}, S.F., 1996, ApJL, 462, L5
  
\bibitem[\protect\citeauthoryear{{Schmalzing}}{1999}]{schmlz99}{Schmalzing},J.,
  PhD Thesis, Ludwig-Maxmilians-Universitat Munchen
  
\bibitem[\protect\citeauthoryear{{Schmalzing} \&
    {Buchert}}{1997}]{Krofton97} {Schmalzing}, J.  \& {Buchert}, T.,
  1997, ApJL, 482, L1

\bibitem[\protect\citeauthoryear{{Schmalzing} et al.}
  {1999}]{schmal99} {Schmalzing}, J., {Buchert}, T., Melott, A.L.,
  Sahni, V., {Sathyaprakash}, B.S.  and {Shandarin}, S.F., 1999, ApJ,
  526, 568.
  
\bibitem[\protect\citeauthoryear{{Shandarin} \& {Zeldovich}}{1983}]
  {shz83} {Shandarin}, S.F. \& {Zeldovich}, Ya.B., 1983, Comments
  Astrophys.  10, 33
  
\bibitem[\protect\citeauthoryear{{Shandarin} \& {Zeldovich}}{1989}]
  {shandzed89} {Shandarin}, S.F. \& {Zeldovich}, Ya.B., 1989, Rev.
  Mod. Phys.,61,185
  
\bibitem[\protect\citeauthoryear{{Shandarin}, {Sheth} \& {Sahni}}{2003}]{shand03}
  {Shandarin}, S.F., {Sheth}, J.V., {Sahni}, V., 2003, In preparation.

\bibitem[\protect\citeauthoryear{{Sheth} et al.}{2003}]{sheth03}
  {Sheth}, J.V., {Sahni}, V., {Shandarin}, S.F. \& {Sathyaprakash}, B.S.,
  2003, MNRAS, 343, 22
                
\bibitem[\protect\citeauthoryear{{Springel} et al.}{1998}]{springtop}
  {Springel}, V. et al. (for the Virgo Consortium), 1998, MNRAS, 298,
  1169
  
\bibitem[\protect\citeauthoryear{{Weinberg}, {Hernquist} \& {Katz}}{2002}]{weinberg02}
  {Weinberg}, D.H., {Hernquist}, L., \& {Katz}, N., 2002, ApJ, 571, 15
  
\bibitem[\protect\citeauthoryear{{Weinberg}}{1988}]{weinberg88}
  {Weinberg}, D. H., 1988, PASP, 100, 1373

\bibitem[\protect\citeauthoryear{{White}, {Efstathiou} \&
    {Frenk}}{1993}]{wef93}{White}, S.D.M., {Efstathiou}, G. and
  {Frenk}, C. S., 1993, MNRAS, 262, 1023

\bibitem[\protect\citeauthoryear{{Yan}, {Madgwick} \& {White}}{2003}]{yan03}
  {Yan}, R., {Madgwick}, D. and {White}, M., 2003, astro-ph/0307248, 
  Submitted to ApJ
  
\bibitem[\protect\citeauthoryear{{Yang}, {Mo}, {van den Bosch} \&
    {Jing}}{2003}]{yang03}{Yang}, X., {Mo}, H. J., {van den Bosch}, F.
  2003, MNRAS, 339, 1057
  
\bibitem[\protect\citeauthoryear{{Yess} \&
    {Shandarin}}{1996}]{ys96} Yess, C. and {Shandarin}, S.F., 1996,
  ApJ, 465, 2.
  
\end{thebibliography}
\end{document}